\documentclass[12pt]{article}
\usepackage{graphics, graphicx, amsmath, subfigure, amsthm, amsfonts, enumerate, epstopdf, booktabs, bm}
\usepackage{times,amsmath,amsfonts,amsthm,amssymb,eucal}
\usepackage{algorithm}
\usepackage{algorithmic}
\usepackage{graphicx,ifthen}
\usepackage{wrapfig}
\usepackage{latexsym}
\usepackage{color}
\usepackage{mathrsfs}
\usepackage{bm}
\usepackage{bbm}
\DeclareGraphicsExtensions{.gif,.pdf,.png,.jpg,.tiff}
\begin{document}
\renewcommand{\baselinestretch}{1.60}
\def\Quote{\begin{quotation}\normalfont\small}
\def\EndQuote{\end{quotation}\rm}
\def\BigHeading{\bfseries\Large}\def\MediumHeading{\bfseries\large}
\def\bct{\begin{center}}
\def\ect{\end{center}}
\font\BigCaps=cmcsc9 scaled \magstep 1
\font\BigSlant=cmsl10    scaled \magstep 1
\def\lbk{\linebreak}
\def\Report{Random-effects Approach to Regression}
\def\Author{}
\pagestyle{myheadings}
\markboth{\Author}{\Report}
\thispagestyle{empty}
\bct{\BigHeading A Random-effects Approach to Regression Involving Many Categorical Predictors and
Their Interactions}
\\\vskip10pt
\BigCaps {Hanmei ${\rm Sun}^{1}$, Jiangshan ${\rm Zhang}^{2}$ and Jiming ${\rm Jiang}^{2}$\lbk
\BigSlant School of Mathematics and Statistics, Shandong Normal University, ${\rm China}^{1}$
and Department of Statistics, University of California, Davis, ${\rm USA}^{2}$}
\ect
\Quote
\vskip-5pt
Linear model prediction with a large number of potential predictors is both statistically and computationally challenging.
The traditional approaches are largely based on shrinkage selection/estimation methods, which are applicable even
when the number of potential predictors is (much) larger than the sample size. A situation of the latter scenario occurs
when the candidate predictors involve many binary indicators corresponding to categories of some categorical predictors
as well as their interactions. We propose an alternative approach to the shrinkage prediction methods in such a case
based on mixed model prediction, which effectively treats combinations of the categorical effects as random effects. We
establish theoretical validity of the proposed method, and demonstrate empirically its advantage over the shrinkage
methods. We also develop measures of uncertainty for the proposed method and evaluate their performance empirically.
A real-data example is considered.

\EndQuote
\section{Introduction}\label{sec:intro}
\hspace{4mm}
Mixed model prediction (MMP; e.g., Jiang and Nguyen 2021, sec. 2.3) has a fairly long history starting with
Henderson's early work in animal breeding (Henderson 1948). The field has since flourished, thanks to its broad
applications in various fields. The traditional fields of applications include genetics, agriculture, education, and
surveys (e.g., Robinson 1991). This is a field where frequentist and Bayesian approaches found common grounds.
Nowadays, new and challenging problems have emerged from such fields as business and health sciences, in
addition to the traditional fields, to which methods of MMP are applicable, or potentially applicable. Many of these
problems occur when interest is at subject level or sub-population level, such as precision medicine (e.g.,
Pennello and Yang 2021) and small area estimation (e.g., Rao and Molina 2015). Besides, linear mixed model is also widely used in longitudinal data analysis, Verbeke and Molenberghs (2000) and Cheng et al. (2010) introduced some guidelines on building mixed models for longitudinal data. On the other hand, The application of linear mixed models is not limited to data with typical structural characteristics, Liu et al. (2007) established a close connection between kernel machine methods and linear mixed models, and all the model parameters can be estimated with the unified linear mixed model framework.

High-dimensionality is among the main features of modern data science. When it comes to regression, it is
desirable to utilize information from a large number of potential predictors. One particular situation, where
such a high-dimensional regression problem may occur, is when the potential predictors under consideration
involve many categorical variables as well as their interactions. In fact, even a few categorical variables with
relatively small numbers of categories can end up with many potential predictors, if interactions are considered.
For example, the ``Bone marrow transplant: children Data Set'' in the UCI Machine Learning Repository was
collected from 187 pediatric patients with 39 attributes, but some attributes describe similar information, such as
donor\_age and donor\_age\_below\_35. A regression analysis is considered with the outcome variable being
the survival time of patients. The predictors involve 6 continuous variables and 8 categorical variables. See
See Table 1 of the supplement for the variable explanations. Among the categorical variables, CMV\_status has 4
categories and HLA\_group\_1 has 7 categories; the rest all have two categories. Suppose that the main interest
is estimating the mean survival times. After removing missing values, there are 166 samples left for the analysis.
However, if we consider the main effects and two-way and three-way interactions among the categorical variables, the the total number of predictors is $6+15+87+263=371$, far exceeding the sample size.

When the number of predictors exceeds the sample size, It is not feasible to fit the regression via the least
squares. The standard practice is to then fit the regression using a shrinkage selection/estimation method, such
as Lasso (Tibshirani 1996), SCAD (Fan and Li 2001), or elastic net (Zou and Hastie 2005). Such a method
amounts to produce shrinkage estimates of the regression coefficients in the sense that a (large) portion of the
coefficients are shrunk to zero, thus achieving variable selection and parameter estimation at the same time.
Once the shrinkage estimates are obtained, the regression function can be estimated via a linear combination
of the nonzero estimated regression coefficients and the corresponding predictors.

The main purpose of the current paper is to propose, and develop, an alternative approach to estimating
the regression mean in such a high-dimensional situation, where a large number of categorical variables are
considered as predictors. The new approach is based on mixed model prediction (MMP; e.g., Jiang and
Nguyen 2021, sec. 2.3). This allows us to reduce the high-dimensional problem to a lower dimensional one
and, more importantly, to focus on characteristics of direct interest.

The method is described in detail in Section 2, followed by a simulated example in Section 3. In Section 4,
we study asymptotic behaviors of the proposed estimators and predictors. In Section 5, we discuss measures
of uncertainty associated with the predictors. More simulation results are presented in Section 6, including
comparison of our new method with Lasso and elastic net, and empirical performance of the proposed
measures of uncertainty. The bone marrow data is revisited in Section 7. Some discussion and concluding
remarks are offered in Section 8. Proofs and technical details are deferred to the supplementary material.
\section{A pseudo MMP approach}\label{sec:PMMP}
\hspace{4mm}
We are going to make some structural change for the part of the regression model involving the categorical
predictors. There may also be continuous predictors, but those remain unchanged. It should be noted that,
although the proposed method is intended for estimation of the mean response, or outcome, it can also be
used for interpretation of the relationship between the outcome variable and the continuous predictors. See
Section 8.

As mentioned, our basic idea is based on MMP. A defining feature of a mixed effects model is random effects.
For prediction under a mixed effects model, MMP is naturally applied (e.g., Jiang and Nguyen 2021, sec. 2.3,
Rao and Molina 2015). However, here we are dealing with a fixed effects model.  Suppose that there are
$N$ samples. The responses, or outcomes, are $y_i, i=1, \ldots, N$. The predictors can be divided into 2
types. Let $x_{i}$ be a $p\times 1$ vector of continuous variables, and $c_{i}=(c_{ij})_{1\leq j\leq q}$ be a
$q\times 1$ vector of categorical variables. which may correspond to the main effects or interactions. The indictor variables, such as $1_{(c_{ij}=k)}, k=1,\dots,C_{j}$, are what we call categorical predictors included in the regression model. Without loss of generality, let $j=1,\dots,q_{1}$ be associated with the main-effects and $j=q_{1}+1,\dots,q$ be with the interactions.
It is need to say, when $c_{ij}$ corresponds to a main effect, the the $j$th categorical variable has  $C_j+1$ categories, denoted by $1, \dots, C_j, C_{j}+1$, where the last category is selected to be the reference category.
For example, there are 4 blood types (A, B, AB, O) but only 3 categorical predictors are included in the regression model for the associated categorical variable, which may correspond to A, B and AB.
When $c_{ij}$ corresponds to the interaction between the categorical variables, its value $k$ is the intersection of the categorical variables.

To illustrate with an example, suppose that the regression involves one continuous variable and three categorical variables, so $p=1, q_1=3$. The categorical variables have 4, 5, and 6 categories, respectively, that is, for the main effect, $C_1=3, C_2=4$ and $C_3=5$. Besides the main effects, if one is to
consider all possible two-way and three-way interactions, we have $q=q_1+3+1=7$. Specifically, if $c_{ij}$
corresponds to the first main-effect, the different categories are $1, 2, 3$; if $c_{ij}$ corresponds to the interaction between the first and second main effects, the different categories are $(1,1), \dots, (1,4), \dots, (3,1), \dots, (3,4)$, hence $C_{j}=3\times4=12$. In total, there are $3+4+5+3\times 4
+3\times 5+4\times 5+3\times 4\times 5=119$ possible indicators of main effects and interactions; in other
words, the total number of categorical predictors is 119.

The underlying model can be expressed as
\begin{equation}
y_i=b_0+x_{i}'b+\sum_{j=1}^{q}\sum_{k=1}^{C_j}a_{jk}1_{(c_{ij}=k)}+\epsilon_{i}, \ \ i=1, \ldots, N.
\label{eq:model_1}
\end{equation}
where $b=(b_{k})_{1\leq k\leq p}$, $b_{k}, 0\leq k\leq p$, $a_{jk}, 1\leq j\leq q, 1\leq k\leq C_{j}$ are
unknown regression coefficients, and $\epsilon_i$'s are i.i.d. regression errors, with mean $0$ and
unknown variance $\sigma^2$. Our main interest is to estimate the regression mean,
\begin{equation}
\theta_{i}=b_0+x_{i}'b+\sum_{j=1}^{q}\sum_{k=1}^{C_j}a_{jk}1_{(c_{ij}=k)},\ \ i=1, \ldots, N.
\label{eq:theta_i}
\end{equation}

Without loss of generality, we can arrange the $N$ samples by the categorical variable categories from
$1$ to $C_j+1$ ($C_j+1$ corresponds to the last reference category), for $j=1, \ldots, q_{1}$ . Specifically,
we first list the samples with $c_{ij}=1$, for all $j=1,\ldots, q_{1}$; then the samples with $c_{ij}=1, j=1,
\ldots, q_{1}-1$ and $c_{iq_{1}}=2$; ......; then the samples with $c_{ij}=1, j=1,\ldots, q_{1}-1$ and
$c_{iq_{1}}=C_{q_{1}}+1$; ......; and finally the samples with $c_{ij}=C_j+1$ for all $j=1,\ldots, q_{1}$. This
way, we can classify the $N$ samples into $K$ groups according the functional value (in terms of the
regression coefficients) of
\begin{eqnarray}
w_{i}=\sum_{j=1}^{q}\sum_{k=1}^{C_j}a_{jk}1_{(c_{ij}=k)},\label{eq:value_i}
\end{eqnarray}
where $K \leq\{\prod_{j=1}^{q_{1}}(C_j+1)\}\wedge N$ [$u\wedge v=\min(u,v)$] is the total number of
different functional values, $w_{i}$, appearing in the samples. For example, in the above illustrative
example, $K\leq(4\times
5\times 6)\wedge 100=100$; however, the actual value of $K$ could be (much) smaller, which is a main
motivation for our proposed method (see below for further discussion).

In practice, the combinations of main effects and interactions appear in the model for practical reasons.
For example, in a medical study, the researchers are interested in the interaction between treatment, a
categorical variables with three categories (placebo, low, high), and sex, a categorical variable with two
categories (female, male), and age, a categorical variable with 9 age groups. There is little interest in this
study about the interaction between sex and age. Thus, the interactions between the sex and age
are not included in the fitted model, and we have no interest in estimating linear combinations involving these
interactions. There is, however, another scenario, in which a linear combination is of interest, but there are
no data associated with the linear combinations. This typically occurs in observational studies rather than in
planned studies. In such a case, the linear combination also does not appear the model (\ref{eq:model_1}).
Although our method does not directly apply to estimating such linear combinations, a modification can
make our method apply. The idea is to include the main effect or interactions involved in such a ``missing
linear combination'' in the $x_{i}$ part (together with the continuous covariates). We can then estimate the
corresponding regression coefficients, and use them to estimate the linear combination, just like what one
typically does in standard regression. See the first paragraph of Section 8 for more discussion.

Denote the $K$ groups by ${\cal G}_1,\ldots, {\cal G}_K$ with $|{\cal G}_{k}|=n_{k}, 1\leq k\leq K$ ($|A|$
denotes the cardinality of set $A$). Note that the data in each group have the same $w_{i}$, which is
the part associated with the categorical predictors in (\ref{eq:model_1}) or (\ref{eq:theta_i}). Denote the
$w_{i}$ by $\alpha_{k}$ for $i\in{\cal G}_{k}, 1\leq k\leq K$. However, the value of $x_{i}$ may be different
for $i\in{\cal G}_{k}$. Let $x_{kl}$ denote the $l$th (vector) value of $x_{i}$ in group ${\cal G}_{k}$, $1\leq
l\leq n_{k}$, $1\leq k\leq K$; similarly for $y_{i}$ and $\epsilon_{i}$. Then, model (\ref{eq:model_1}) can be
expressed in a different way:
\begin{eqnarray}
y_{kl}=b_{0}+x_{kl}'b+\alpha_{k}+\epsilon_{kl},\;\;l=1,\dots,n_{k},\;k=1,\dots,K.\label{eq:model_2}
\end{eqnarray}
Model (\ref{eq:model_2}) can be expressed in the standard matrix expression of a linear mixed
model (LMM; e.g., Jiang and Nguyen 2021, sec. 1.1). Let $y_{[k]}=(y_{kl})_{1\leq l\leq n_{k}}$
($n_{k}\times 1$) and define $\epsilon_{k}$ similarly; let $X_{[k]}=[(1\;x_{kl}')]_{1\leq l\leq n_{k}}$
[$n_{k}\times(p+1)$ matrix]. Then, let $y=(y_{[k]})_{1\leq k\leq K}$, $\epsilon=(\epsilon_{k})_{1\leq
k\leq K}$, $X=(X_{[k]})_{1\leq k\leq K}$ (stacking the vectors or matrices), and $Z={\rm diag}(
1_{n_{k}}, 1\leq k\leq K)$, where $1_{n}$ denotes the $n\times 1$ vector of $1$s and ${\rm
diag}(A_{k}, 1\leq k\leq K)$ the block-diagonal matrix with $A_{1},\dots,A_{K}$ on the diagonal.
Finally, define $\beta=(b_{0}, b')'$ and $\alpha=(\alpha_{k})_{1\leq k\leq K}$ ($K\times 1$ vector).
Then, model (\ref{eq:model_2}) can be expressed as
\begin{eqnarray}
y=X\beta+Z\alpha+\epsilon.\label{eq:model_mat}
\end{eqnarray}
Note that $N=\sum_{k=1}^{K}n_{k}$ and $y$ is an $N\times 1$ vector. The order of the
components of vectors, and rows of matrices, in (\ref{eq:model_mat}) can be arranged to be the
same as in (\ref{eq:model_1}), following the ordering described below (\ref{eq:theta_i}), after
removing the empty cells, so that we have, component-wisely,
\begin{eqnarray}
y_{i}=\theta_{i}+\epsilon_{i},\;\;\theta_{i}=b_{0}+x_{i}'b+z_{i}'\alpha,\;\;i=1,\dots,N,
\label{eq:theta_i_new}
\end{eqnarray}
where $z_{i}'$ is the $i$th row of $Z$.  Comparing (\ref{eq:theta_i_new}) with
(\ref{eq:model_1})--(\ref{eq:value_i}), we see the only difference is that $w_{i}$ is replaced by
$z_{i}'\alpha$, which is equal to $\alpha_{k}$ for $i\in{\cal G}_{k}$.

The good news is that (\ref{eq:model_mat}), or (\ref{eq:theta_i_new}), is in the standard LMM
formation, even though there is actually no random effect; this is right---we have ``created''
some ``random effects'' just so that we can apply MMP. Under the assumption that $\alpha\sim
N(0,GI_{m})$, $\epsilon\sim N(0,RI_{N})$, and $\alpha$ is independent with $\epsilon$, where
$G, R$ are unknown variance components and $I_{n}$ denotes the $n$-dimensional identity
matrix, the empirical best linear unbiased predictor (EBLUP; e.g., Jiang and Nguyen 2021, sec.
2.3) of $\theta_{i}$ is given by
\begin{eqnarray}
\hat{\theta}_{i}=\hat{b}_{0}+x_{i}'\hat{b}+z_{i}'\hat{\alpha},\;\;z_{i}'\hat{\alpha}=
\frac{\hat{h}n_{k}}{1+\hat{h}n_{k}}(\bar{y}_{k\cdot}-\hat{b}_{0}-\bar{x}_{k\cdot}'\hat{b}),
\label{eq:eblup}
\end{eqnarray}
where $\hat{\beta}=(\hat{b}_{0},\hat{b}')'=(X'\hat{H}^{-1}X)^{-1}X'\hat{H}^{-1}y$ with $\hat{H}
=I_{N}+\hat{h}ZZ'$ and $\hat{h}=\hat{G}/\hat{R}$, $k$ being the group index such that
$z_{i}'\alpha=\alpha_{k}$, $\bar{y}_{k\cdot}=n_{k}^{-1}\sum_{l=1}^{n_{k}}y_{kl}$, and
$\bar{x}_{k\cdot}=n_{k}^{-1}\sum_{l=1}^{n_{k}}x_{kl}$. Here, $\hat{h}$ is an estimator of $h=
G/R$ with $\hat{G}, \hat{R}$ being the estimators of $G, R$, respectively; for example,
$\hat{G}, \hat{R}$ may be the maximum likelihood (ML; e.g., Jiang and Nguyen 2021,
sec. 1.3.1) estimators of $G, R$, respectively, under the above LMM assumption (including the
distributional assumption about $\alpha, \epsilon$). Furthermore, $\hat{\beta}$ is the empirical
best linear unbiased estimator (EBLUE) of $\beta$ under the LMM assumption.

Of course, there are no real random effects, as noted earlier, and all of these distributional
assumptions imposed on $\alpha$ are ``fake''. Nevertheless, the EBLUP targets directly the
characteristic of interest, $\theta_{i}$ in (\ref{eq:theta_i}). Note that the total number of
$\alpha_{k}$'s associated with $\theta_{i}$, $K$, is guaranteed less, and possibly much
less, than the sample size, $N$. In contrast, if one were to estimate $\theta_{i}$ via the
least squares (in case it is feasible) or shrinkage selection/estimation methods, one would have
to first estimate all of the regression coefficients associated with the predictors, continuous
or categorical, the total number of which could be much larger than the sample size. In fact,
many of these coefficients may only appear a few times in (\ref{eq:model_1}) with the data
so, intuitively, there is not sufficient information in estimating them individually. More
importantly, if the ultimate goal is to estimate $\theta_{i}$, why not targeting it directly,
rather than going around the seemingly inefficient, and possibly expensive, route of first
estimating the numerous regression coefficients? At least from this point of view, the new
method introduced above, which we call pseudo MMP (PMMP), seems to be more reasonable,
especially if we can justify it theoretically.

That is, of course, a big if at this point, and a main purpose of the rest of the paper. But before
providing a theoretical justification, let us first demonstrate, empirically, the performance of
EBLUP in comparison with the shrinkage selection/estimation method with a simulated example.
\section{A simulated example}\label{sec:sim1}
\hspace{4mm}
The example follows the lines of the illustrative example in Section 2. We consider a
scenario similar to Zou and Hastie (2005), with $b_0=1, b_1=2$ and the 119 regression
coefficients for the categorical predictors given by
$a=(\underbrace{2, \ldots, 2}_{29}, \underbrace{0,\ldots,0}_{30}, \underbrace{2,\ldots, 2}_{30},
\underbrace{0,\ldots, 0}_{30})$.
The sample size is $N=30$.

The continuous and categorical predictors are generated following the specifications introduced in
Section 6.1. According to the different values of categorical predictors, we can arrange the 30 samples into
$K$ groups. Here, for the generated data, after removing the empty groups, $K$ is 26. In other words, if
we want to fit the data by the linear mixed model, \eqref{eq:eblup}, there are two fixed effects, $b_0$ and
$b_1$, and $K=26$ group-specific random effects in the model. Thus, the number of random effects is
much less than 121, which is the total number of the regression coefficients. The random effects are
then formulated by applying the procedure described in Section 2. For example, the first random effect is
$\alpha_1=a_{22}1_{(c_{i2=2})}+a_{32}1_{(c_{i3=2})}+a_{622}1_{(c_{i2=2})}1_{(c_{i3=2})}$; the second
random effect is $\alpha_2=a_{22}1_{(c_{i2=2})}+a_{34}1_{(c_{i3=4})}+a_{624}1_{(c_{i2=2})}1_{(c_{i3=4})}$,
and so on [see (\ref{eq:model_1}) for notation; a more specific expression is given in (\ref{eq:sim_model})].

We compare the averaged squared error (ASE) of Lasso, elastic net, and PMMP, for estimating all
regression means. The ASE is defined as
\begin{eqnarray}
{\rm ASE}=\frac{1}{N}\sum_{i=1}^{N}(\tilde{\theta}_{i}-\theta_{i})^{2},\label{eq:ase}
\end{eqnarray}
where $\tilde{\theta}_{i}$ may correspond to Lasso, elastic net, or PMMP (i.e., $\hat{\theta}_{i}$).
The Lasso and elastic net are computed using the {\bf glmnet} package, with the selection of
the $\alpha$-parameter for elastic net chosen from $0, 0.1, 0.2, \dots, 1$ using 10-fold
cross-validation. We carried out $N_{\rm sim}=200$ simulation runs. Boxplots of the $200$
ASEs are presented in Figure 1. It appears that PMMP is a clear winner in terms of the ASE.

\begin{figure}[t!]
  \centering
  \includegraphics[width=0.75\textwidth,height=6cm]{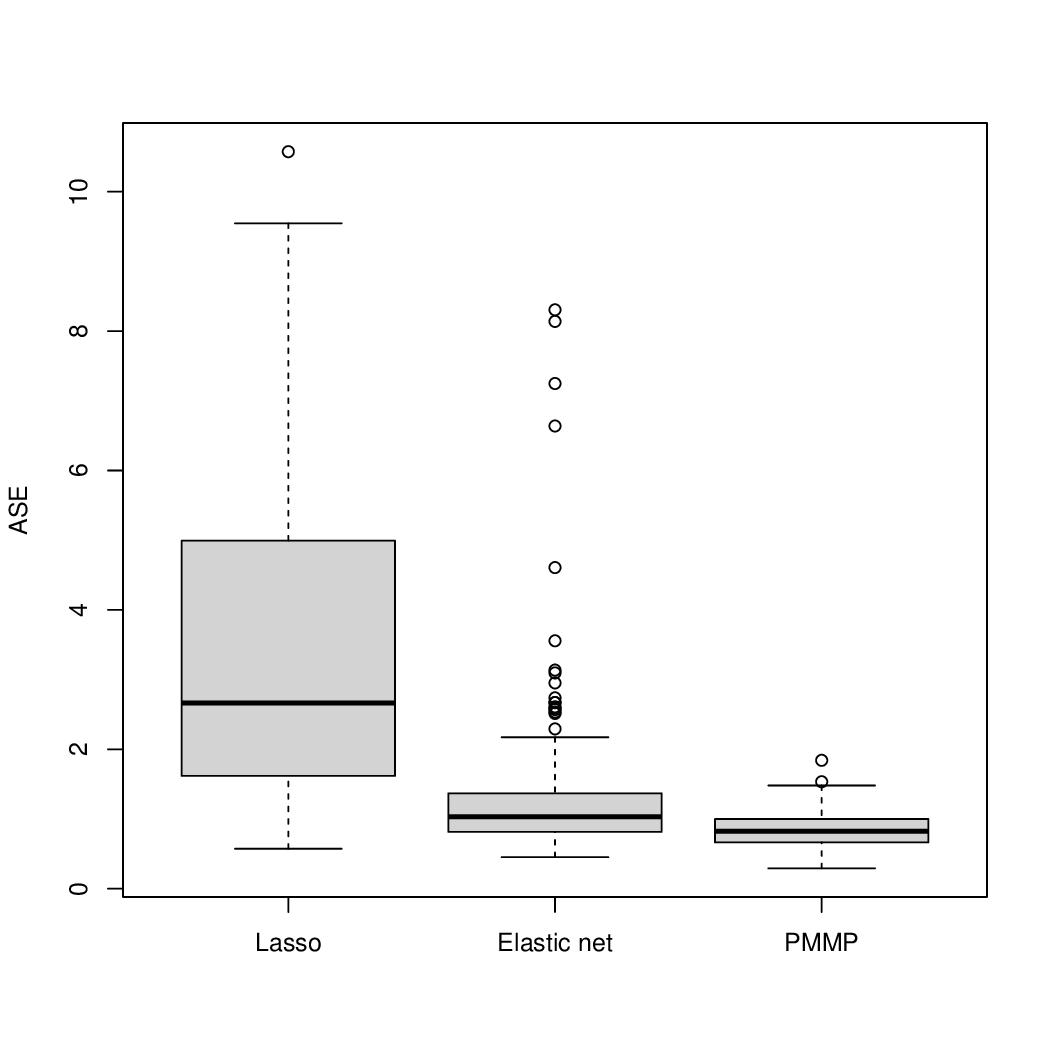}
  \caption{Boxplots of ASEs (Sparse Scenario; $N=30$, $N_{\rm sim}=200$)}
\end{figure}
\section{Asymptotic theory}
\label{sec:2nd}
\subsection{Convergence of pseudo MLEs}
\label{sec:2nd-1}
\hspace{4mm}
In this subsection, we show that, under regularity conditions, the pseudo MLEs converge in
probability to certain limits with reasonable interpretations. By Jiang and Nguyen (2021, sec.
1.3.1), the log-likelihood function under the LMM assumption, multiplied by $-2$, can be
expressed as
\begin{eqnarray}
 Q&=&c+N\log R+\log|H|\nonumber\\
 &&+\frac{1}{R}(y-1_N b_0-X_{1}b)'H^{-1}(y-1_N b_0-X_{1}b),
 \label{eq:Q_def}
\end{eqnarray}
where $|H|=\prod_{k=1}^{K}(1+n_{k}h)$, $H^{-1}={\rm diag}\{I_{n_k}-h(1+n_{k}h)^{-1}1_{n_k}1_{n_k}'\}$,
$X_{1}=(x_{i}')_{1\leq i\leq N}$, which is $N\times p$, and $c=N\log(2\pi)$. By differentiating $Q$ with
respect to the parameters, $b_{0}, b, R$ and $h$, we have the ML equations:
\begin{eqnarray}
 &&\frac{\partial Q}{\partial b_0}=\frac{2}{R}1_{N}'H^{-1}(y-1_N b_0-X_{1}b)=0,\label{eq:eq_b0}\\
 &&\frac{\partial Q}{\partial b}=\frac{2}{R}X_{1}'H^{-1}(y-1_N b_0-X_{1}b)=0,\label{eq:eq_b}\\
 &&\frac{\partial Q}{\partial R}=\frac{N}{R}-\frac{1}{R^2}(y-1_N b_0-X_{1}b)'H^{-1}(y-1_N b_0-X_{1}b)\nonumber\\
 &&=0,
 \label{eq:eq_R}\\
 &&\frac{\partial Q}{\partial h}=\sum_{k=1}^{K}\left[\frac{n_k}{1+hn_{k}}-\frac{1}{R}\left\{\frac{n_{k}}{1+
 hn_{k}}(\bar{y}_{k\cdot}-b_{0}-\bar{x}_{k\cdot}'b)\right\}^{2}\right]\nonumber\\
 &&=0.\label{eq:eq_h}
\end{eqnarray}
The pseudo MLEs, $\hat{b}_{0}, \hat{b}, \hat{R}$ and $\hat{h}$, are solution to the ML equations,
(\ref{eq:eq_b0})--(\ref{eq:eq_h}). To ensure good asymptotic behavior, the estimators are obtained via the
following procedure. Let $h_{N}$ be a sequence of constants that satisfy assumption {\it A3} below. The
sequence $h_{N}$ is used to regularize the solution to the ML equations, following the below arguments
and procedures:\\
(a) It can be shown that (\ref{eq:eq_b0})--(\ref{eq:eq_R}), with $h=h_{N}$, have a closed-form solution,
say, $\tilde{b}_{0}, \tilde{b}, \tilde{R}$, that satisfy the conclusions of (i)--(iii) of Theorem 1 below, with
$\hat{b}_{0}, \hat{b}, \hat{R}$ replaced by $\tilde{b}_{0}, \tilde{b},\tilde{R}$, respectively [see the proof of
(i)--(iii) of Theorem 1].\\
(b) It can then be shown [see the proof of (iv) of Theorem 1] that, with probability tending to one,
(\ref{eq:eq_h}), with $b_{0}, b, R$ replaced by $\tilde{b}_{0}, \tilde{b}, \tilde{R}$, respectively, has a solution,
say, $\tilde{h}$, that satisfies the conclusion of (iv) of Theorem 1, with $\hat{h}$ replaced by $\tilde{h}$.\\
(c) We then solve (\ref{eq:eq_b0})--(\ref{eq:eq_R}), again but this time with $h=\tilde{h}\vee h_{N}$. Once
again, the equations have a closed-form solution, denoted by $\hat{b}_{0}, \hat{b}, \hat{R}$, respectively,
that satisfy the conclusions of (i)--(iii) of Theorem 1.\\
(d) Finally, we solve (\ref{eq:eq_h}), again but this time with $b_{0}, b, R$ replaced by $\hat{b}_{0}, \hat{b},
\hat{R}$, respective. Again, with probability tending to one, the equation has a solution, denoted
by $\hat{h}$, that satisfies conclusion (iv) of Theorem 1.\\
(e) For computing the EBLUPs, (\ref{eq:eblup}), replace the $\hat{h}$ in (d) by $\hat{h}\vee h_{N}$. The
result is still denoted by $\hat{h}$ for notation simplicity.

We assume the following regularity conditions.\\
{\it A1.} The true regression coefficients $b_{k}, 0\leq k\leq p$ and $a_{jk}, 1\leq j\leq q, 1\leq k\leq
C_{j}$ in (\ref{eq:model_1}) are bounded, and $\sigma^{2}\in(0,\infty)$.\\
{\it A2.} All the elements of $X_{1}$ are bounded, and
\begin{eqnarray}
\liminf\lambda_{\min}\left[\frac{1}{N}\sum_{k=1}^{K}\sum_{i\in{\cal G}_{k}}(x_{i}-\bar{x}_{k\cdot})(x_{i}
-\bar{x}_{k\cdot})'\right]>0,\label{eq:eigen_lb}
\end{eqnarray}
where $\lambda_{\min}$ denotes the smallest eigenvalue.\\
{\it A3.} $h_{N}\rightarrow 0$ and $h_{N}n_{*}\rightarrow\infty$, where $n_{*}=\min_{1\leq k\leq K}n_{k}$.

Note that the assumptions have nothing to do with the working LMM (\ref{eq:model_mat}); in other words,
the assumptions are regarding the true data generating model (\ref{eq:model_1}). Specifically, assumption {\it
A1} is clearly reasonable. Assumption {\it A2} has some implication about the relative sizes of $K$ and $N$.
For example, assuming that the continuous variables, $x_{i}$, are bounded, and there are a bounded number
of different $x_{i}$'s in each group $k$, $1\leq k\leq K$. Then, {\it A2} suggests that the relative sizes of $K$
and $N$ are comparable [because, if $K/N\rightarrow 0$, the left side of (\ref{eq:eigen_lb}) would go to zero].(There seem to be some issues with the expression here. In our previous discussion, we have already established the upper limit of K, indicating that K does not tend to infinity as N increases. Of course, this does not lead to the left side of the equation being 0, because the number of terms in the summation is related to N.)
Assumption {\it A3} is regarding $h_{N}$, a constant sequence used to regularize $\hat{h}$. Basically, {\it A3}
means that $h_N$ goes to zero but not too fast so that $h_{N}^{-1}=o(n_{*})$. For example, assuming
$n_{*}\rightarrow\infty$, one may choose $h_{N}=\delta/\sqrt{n_{*}}$, where $\delta$ is any given (small)
positive constant. Then, clearly, assumption {\it A3} holds.

{\bf Theorem 1.} Under assumptions {\it A1}--{\it A3}, the following hold:
(i) $\hat{b}_{0}=b_{0}+\bar{\alpha}+o_{\rm P}(1)$, where $\bar{\alpha}=K^{-1}\sum_{k=1}^{K}\alpha_{k}$;
in particular, if also $K\rightarrow\infty$ and $\lim_{K\rightarrow\infty}\bar{\alpha}=\alpha_{0}\in{\mathcal
R}$ (the space of real numbers), then, we have $\hat{b}_{0}\stackrel{\rm P}{\longrightarrow}b_{0}
+\alpha_{0}$.
(ii) $\hat{b}\stackrel{\rm P}{\longrightarrow}b$.
(iii) $\hat{R}\stackrel{\rm P}{\longrightarrow}R
=\sigma^{2}$. (iv) With probability tending to one, equation (\ref{eq:eq_h}) has a solution, $\hat{h}$,
satisfying
$$\hat{h}=\frac{1}{RK}\sum_{k=1}^{K}(\alpha_{k}-\bar{\alpha})^{2}+o_{\rm P}(1)=h_{K}+o_{\rm P}(1),$$
with $h_{K}$ defined in an obvious way, provided that $G_{K}=K^{-1}\sum_{k=1}^{K}(\alpha_{k}-\bar{\alpha})^{2}$
is bounded, and bounded away from zero; in particular, if also $K\rightarrow\infty$ and
$\lim_{K\rightarrow\infty}G_{K}=G\in(0,\infty)$, then, we have $\hat{h}\stackrel{\rm P}{\longrightarrow}h=G/R$.

The proof of Theorem 1 is given in Section 1 of the supplementary material. It is seen that $\hat{b}$ and
$\hat{R}$ are consistent estimators, while $\hat{b}_{0}$ is not consistent, unless $\bar{\alpha}\rightarrow 0$.
Note that there is no $h$ in the real world so we do not talk about consistency of $\hat{h}$; however, it does
converge in probability to something that is reasonable. It can be seen from Theorem 1 that $h$ is $G/R$,
where $R=\sigma^{2}$, the variance of the regression errors, and $G$ is the limit of the sample variance of the
``working'' random effects corresponding to the linear combinations of the categorical effects, assuming that the
limit exists. Thus, $h$ can be interpreted as the signal to noise ratio corresponding to the categorical part of the
mean function.
\subsection{Consistency and $L^{2}$ convergence of pseudo EBLUPs}
\label{sec:2nd-2}
\hspace{4mm}
The convergency of the MLEs, $\hat{b}_0, \hat{b}, \hat{G}$ and $\hat{R}$, leads to consistency of the pseudo
EBLUPs, (\ref{eq:eblup}). Note that
$\bar{y}_{k\cdot}=n_{k}^{-1}\sum_{l=1}^{n_k}(b_0+x_{kl}'b+\alpha_{k}+\epsilon_{kl})=b_0+\bar{x}_{k\cdot}'b
+\alpha_{k}+\bar{\epsilon}_{k\cdot}$. It can then be shown, using (i), (ii) of Theorem 1 and the property of
$\hat{h}$, that for $i\in{\cal G}_{k}$,
\begin{eqnarray}
\hat{\theta}_{i}-\theta_{i}&=&\hat{b}_{0}+x_{i}'\hat{b}+\frac{\hat{h}n_{k}}{1+\hat{h}n_{k}}\{b_{0}-\hat{b}_{0}
+\bar{x}_{k\cdot}'(b-\hat{b})+\alpha_{k}+\bar{\epsilon}_{k\cdot}\}\nonumber\\
&&-b_{0}-x_{i}'b-\alpha_{k}\nonumber\\
&=&\hat{b}_{0}-b_{0}+x_{i}'(\hat{b}-b)-\alpha_{k}\nonumber\\
&&+\left(1-\frac{1}{1+\hat{h}n_{k}}\right)\{b_{0}-\hat{b}_{0}+\bar{x}_{k\cdot}'(b-\hat{b})+\alpha_{k}
+\bar{\epsilon}_{k\cdot}\}\nonumber\\
&=&\hat{b}_{0}-b_{0}+x_{i}'(\hat{b}-b)-\alpha_{k}+b_{0}-\hat{b}_{0}+\bar{x}_{k\cdot}'(b-\hat{b})+\alpha_{k}
+\bar{\epsilon}_{k\cdot}\nonumber\\
&&+\frac{O_{\rm P}(1)}{1+\hat{h}n_{k}}\nonumber\\
&=&(x_{i}-\bar{x}_{k\cdot})'(\hat{b}-b)+\bar{\epsilon}_{k\cdot}+\frac{O_{\rm P}(1)}{1+\hat{h}n_{k}}\nonumber\\
&=&o_{\rm P}(1).\label{eq:eblup_dif}
\end{eqnarray}
The consistency result can be strengthened to convergence in $L^{2}$, as follows.

{\bf Theorem 2.} Under the conditions of Theorem 1, we have, for every $1\leq k\leq K$ and $i\in{\cal G}_{k}$,
$\hat{\theta}_{i}-\theta_{i}=o_{\rm P}(1)$; in fact, we have ${\rm E}(\hat{\theta}_{i}-\theta_{i})^{2}=o(1)$.

The proof of Theorem 2 is given in Section 2 of the supplementary material. From the proof it can be seen
that, for $i\in{\cal G}_{k}$,  the order of
\begin{eqnarray}
{\rm MSE}(\hat{\theta}_{i})\equiv{\rm E}(\hat{\theta}_{i}-\theta_{i})^{2}
\end{eqnarray}
is $O(n_{*}^{-1})$. If $n_{*}$ is not very large, the MSE can still be significant. In the next section, we develop
a method to estimate the MSE.
\section{Measure of uncertainty}
\label{sec:3rd}
\hspace{4mm}
We assume the conditions of Theorem 1 hold and $K\rightarrow\infty$ and $\lim_{K\rightarrow}G_{K}=
G\in(0,\infty)$. By (\ref{eq:eblup_dif}), it can be seen that, for $i\in{\cal G}_{k}$, we have
\begin{eqnarray}
\hat{\theta}_{i}-\theta_{i}&=&\left(x_{i}-\frac{\hat{h}n_{k}}{1+\hat{h}n_{k}}\bar{x}_{k\cdot}\right)'(\hat{b}-b)
+\frac{\hat{b}_{0}-b_{0}-\bar{\alpha}}{1+\hat{h}n_{k}}\nonumber\\
&&-\frac{\alpha_{k}-\bar{\alpha}}{1+\hat{h}n_{k}}+\frac{\hat{h}n_{k}}{1+\hat{h}n_{k}}\bar{\epsilon}_{k\cdot}.
\label{eq:mse_1}
\end{eqnarray}

By the proof of Theorem 1 [see (1), (11) in the supplement], we have
\begin{eqnarray}
\hat{b}-b&=&(X_{1}'\hat{H}^{-1}X_{1})^{-1}X_{1}'\hat{H}^{-1}(Z\alpha-\bar{\alpha}1_{N})+(X_{1}'\hat{H}^{-1}
X_{1})^{-1}X_{1}'\hat{H}^{-1}\epsilon\nonumber\\
&&-(X_{1}'\hat{H}^{-1}X_{1})^{-1}X_{1}'\hat{H}^{-1}1_{N}(\hat{b}_{0}-b_{0}-\bar{\alpha}),\label{eq:mse_2}\\
\hat{b}_{0}-b_{0}-\bar{\alpha}&=&\frac{1_{N}'\hat{H}^{-1}(I-M_{1})(Z\alpha-\bar{\alpha}1_{N})}{1_{N}'
\hat{H}^{-1}(I-M_{1})1_{N}}+\frac{1_{N}'\hat{H}^{-1}(I-M_{1})\epsilon}{1_{N}'\hat{H}^{-1}(I-M_{1})1_{N}}.
\label{eq:mse_3}
\end{eqnarray}
Let $J_{N}=1_{N}1_{N}'$. Hereafter, lot denotes a term that is of lower order compered to the terms that are
present (lot for ``lower-order term''). Combining (\ref{eq:mse_2}) and (\ref{eq:mse_3}), it can be shown that
\begin{eqnarray}
&&\hat{b}-b=(X_{1}'\hat{H}^{-1}X_{1})^{-1}X_{1}'\hat{H}^{-1}\left\{I_{N}-\frac{J_{N}\hat{H}^{-1}(I-M_{1})}{1_{N}'
\hat{H}^{-1}(I-M_{1})1_{N}}\right\}(Z\alpha-\bar{\alpha}1_{N}+\epsilon)\nonumber\\
&&=\hat{W}(Z\alpha-\bar{\alpha}1_{N}+\epsilon)=W(Z\alpha-\bar{\alpha}1_{N}+\epsilon)+{\rm lot},
\label{eq:mse_4}
\end{eqnarray}
$\hat{W}$ defined in an obvious way and $W=\hat{W}$ with $\hat{h}$ replaced by $h=G/R$.

We need a more explicit expression of $W$. The following expression can be derived (see Section 3 of the
supplementary material):
\begin{eqnarray}
W=W_{1}\left(I_{N}-\frac{W_{2}}{d}\right),\label{eq:W_exp}
\end{eqnarray}
where $d=d_{1}-d_{2}$ with $d_{1}=\sum_{k=1}^{K}n_{k}(1+hn_{k})^{-1}$ and
$$d_{2}=\left(\sum_{k=1}^{K}\frac{n_{k}\bar{x}_{k\cdot}'}{1+hn_{k}}\right)\left(\sum_{k=1}^{K}x_{[k]}'A_{k}x_{[k]}\right)^{-1}\left(\sum_{k=1}^{K}\frac{n_{k}\bar{x}_{k\cdot}}{1+hn_{k}}\right),$$
$$W_{1}=\left(\sum_{k=1}^{K}x_{[k]}'A_{k}x_{[k]}\right)^{-1}(x_{[k]}'A_{k})_{1\leq k\leq K}',$$
and $W_{2}=1_{N}W_{3}$, $W_{3}=U_{21}-U_{22}$ with $U_{21}=[(1+hn_{k})^{-1}1_{n_{k}}]_{1\leq k\leq K}'$ and
$$U_{22}=\left(\sum_{k=1}^{K}\frac{n_{k}\bar{x}_{k\cdot}'}{1+hn_{k}}\right)\left(\sum_{k=1}^{K}x_{[k]}'A_{k}x_{[k]}
\right)^{-1}(x_{[k]}'A_{k})_{1\leq k\leq K}'$$
with $A_{k}=I_{n_{k}}-h(1+hn_{k})^{-1}J_{n_{k}}$ (recall $J_{n}=1_{n}1_{n}'$).

Note that, with the above notation, we can now express (\ref{eq:mse_3}) as
\begin{eqnarray}
\hat{b}_{0}-b_{0}-\bar{\alpha}&=&\left(\frac{W_{3}}{d}\right)(Z\alpha-\bar{\alpha}1_{N}+\epsilon)+{\rm lot}.
\label{eq:mse_5}
\end{eqnarray}

Now define $w_{k}'$ as the $1\times N$ vector, whose first $n_{1}+\cdots+n_{k-1}$ components and last
$n_{k+1}+\cdots+n_{K}$ components are $0$, and middle $n_{k}$ components are $n_{k}^{-1}1_{n_{k}}'$,
that is,
$w_{i}'=(\underbrace{0, \ldots, 0}_{n_{1}+\cdots+n_{k-1}}, n_{k}^{-1}1_{n_{k}}', \underbrace{0,\ldots,
0}_{n_{k+1}+\cdots+n_{K}})$.
Then, it is easy to verify that $\alpha_{k}-\bar{\alpha}=w_{k}'(Z\alpha-\bar{\alpha}1_{N})$ and
$\bar{\epsilon}_{k\cdot}=w_{k}'\epsilon$. Thus, combined with (\ref{eq:mse_1}), (\ref{eq:mse_4}) and
(\ref{eq:mse_5}), we have, for $i\in{\cal G}_{k}$, $\hat{\theta}_{i}-\theta_{i}=$
\begin{eqnarray}
&&\left\{\left(x_{i}-\frac{hn_{k}}{1+hn_{k}}\bar{x}_{k\cdot}\right)'W+\frac{W_{3}}{(1+hn_{k})d}-\frac{w_{k}'}{1
+hn_{k}}\right\}(Z\alpha-\bar{\alpha}1_{N})\nonumber\\
&&+\left\{\left(x_{i}-\frac{hn_{k}}{1+hn_{k}}\bar{x}_{k\cdot}\right)'W+\frac{W_{3}}{(1+hn_{k})d}
+\frac{hn_{k}}{1+hn_{k}}w_{k}'\right\}\epsilon\nonumber\\
&&+{\rm lot}.\label{eq:mse_6}
\end{eqnarray}
This leads to the following expression of the MSE: ${\rm MSE}(\hat{\theta}_{i})\approx$
\begin{eqnarray}
&&\left[\left\{\left(x_{i}-\frac{hn_{k}}{1+hn_{k}}\bar{x}_{k\cdot}\right)'W
+\frac{W_{3}}{(1+hn_{k})d}-\frac{w_{k}'}{1+hn_{k}}\right\}(Z\alpha-\bar{\alpha}1_{N})\right]^{2}\nonumber\\
&&+\sigma^{2}\left|\left(x_{i}-\frac{hn_{k}}{1+hn_{k}}\bar{x}_{k\cdot}\right)'W+\frac{W_{3}}{(1+hn_{k})d}
+\frac{hn_{k}}{1+hn_{k}}w_{k}'\right|^{2},\label{eq:mse_7}
\end{eqnarray}
where $|v|=(\sum_{i=1}^{N}v_{i}^{2})^{1/2}$ denotes the Euclidean norm of $v=(v_{i})_{1\leq i\leq N}$.

For $i\in{\cal G}_{k}$, an estimator of the MSE, denoted by $\widehat{\rm MSE}(\hat{\theta}_{i})$, is obtained by
the right side of (\ref{eq:mse_7}) with $\sigma^{2}, h$ replaced by $\hat{R}, \hat{h}$, respectively; as for the
$\alpha$, which also appears on the right side of (\ref{eq:mse_7}), it is replaced by its empirical best linear
unbiased predictor (EBLUP; e.g., Jiang and Nguyen 2021, sec. 2.3), given by (\ref{eq:eblup}), that is,
\begin{eqnarray}
\hat{\alpha}_{k}=\frac{\hat{h}n_{k}}{1+\hat{h}n_{k}}(\bar{y}_{k\cdot}-\hat{b}_{0}-\bar{x}_{k\cdot}'\hat{b}),\;\;
1\leq k\leq K.\label{eq:alpha_ihat}
\end{eqnarray}
\section{More simulation studies}
\label{sec:4th}
\hspace{4mm}
We carried out a series of simulation studies on finite-sample performance of the PMMP as well as the
proposed MSE estimator. In particular, we made comparison with the existing shrinkage methods in
estimating the regression means in our simulation study.
\subsection{Performance of PMMP}
\label{sec:4th-1}
\hspace{4mm}
A simulation study was presented in Section 1, in which we made same-data comparisons of the
performance of PMMP with Lasso and elastic net. The simulation was under a ``sparse" scenario. In this
subsection, we make same-data comparisons of PMMP with the those shrinkage methods under a
``dense'' scenario of simulation study. The data are generated under the same model used as an
illustrative example in Section 3, expressed as
\begin{eqnarray}
  y_i&=&b_0+b_{1}x_{i}+\sum_{j=2}^{4}a_{1j}1_{(c_{i1}=j)}+\sum_{j=2}^{5}a_{2j}1_{(c_{i2}=j)}
  +\sum_{j=2}^{6}a_{3j}1_{(c_{i3}=j)}\nonumber\\
  &&+\sum_{j=2}^{4}\sum_{k=2}^{5}a_{4jk}1_{(c_{i1}=j)}1_{(c_{i2}=k)}
  +\sum_{j=2}^{4}\sum_{k=2}^{6}a_{5jk}1_{(c_{i1}=j)}1_{(c_{i3}=k)}\nonumber\\
  &&+\sum_{j=2}^{5}\sum_{k=2}^{6}a_{6jk}1_{(c_{i2}=j)}1_{(c_{i3}=k)}\nonumber\\
  &&+\sum_{j=2}^{4}\sum_{k=2}^{5}\sum_{l=2}^{6}a_{7jkl}1_{(c_{i1}=j)}1_{(c_{i2}=k)}1_{(c_{i3}=l)}
  +\epsilon_i,\label{eq:sim_model}
  \end{eqnarray}
  $i=1, \ldots, N$. Furthermore, we have the following specifications:\\
(1) $x_{i}$ is a continuous predictor, whose values are generated form $N(0,1)$.\\
(2) $c_{ir}, r=1, 2, 3$ are main-effect categorical predictors. Specifically,

(2-1) $c_{i1}$ has 4 categories, denoted by 1, 2, 3, 4; the values of $c_{i1}$ are generated such that ${\rm
P}(c_{i1}=1)={\rm P}(c_{i1}=4)=0.2$, and  ${\rm P}(c_{i1}=2)={\rm P}(c_{i1}=3)=0.3$.

(2-2) $c_{i2}$ has 5 categories, denoted by $1, \ldots, 5$; the values of $c_{i2}$ are generated such that ${\rm
P}(c_{i2}=1)={\rm P}(c_{ic2}=5)=1/12$, ${\rm P}(c_{i2}=2)={\rm P}(c_{i2}=4)=1/4$, and ${\rm P}(c_{i2}=3)=1/3$.

(2-3) $c_{i3}$ has 6 categories, denoted by $1, \ldots, 6$; the values of $c_{i3}$ are generated such that ${\rm
P}(c_{i3}=1)={\rm P}(c_{i3}=6)=1/12$, ${\rm P}(c_{i3}=2)={\rm P}(c_{i3}=5)=1/6$, and ${\rm P}(c_{i3}=3)={\rm
P}(c_{i3}=4)=1/4$.\\
(3) The first line in (\ref{eq:sim_model}) corresponds to the main effects, the second and third lines the
two-way interactions, and the fourth line the three-way interactions. As a result, there are a total of $1+1+3+4+5+3\times4+3\times5+4\times5+3\times4\times5=121$ predictors.\\
(4) The errors $\epsilon_i$ are generated from the $N(0,\sigma^2)$ distribution with $\sigma=1$. We set $b_0=1,
b_1=2$; the other 119 categorical coefficients are generated from ${\rm Uniform}(0,1)$.\\
(5) We consider three different sample sizes: $N=30,50,100$.

For Lasso and Elastic net, we use the those shrinkage methods to estimate the 121 unknown regression coefficients,
then predict $\theta_i$ for every $i$. For PMMP, we use our random-effects approach. Specifically, based on the
value of $c_{ir}, r=1, 2, 3$, divides the data into no more than 120 groups.  A LMM is fitted and the pseudo EBLUPs
of $\theta_i$ are obtained via (\ref{eq:eblup}).

Once again, we use the ASE, (\ref{eq:ase}) as a performance measure. Boxplots of the ASEs, based on 200
simulation runs, are presented in Figure 2.

\begin{figure}[t!]
  \subfigure[$N=30$]{
  \includegraphics[width=0.3\textwidth]{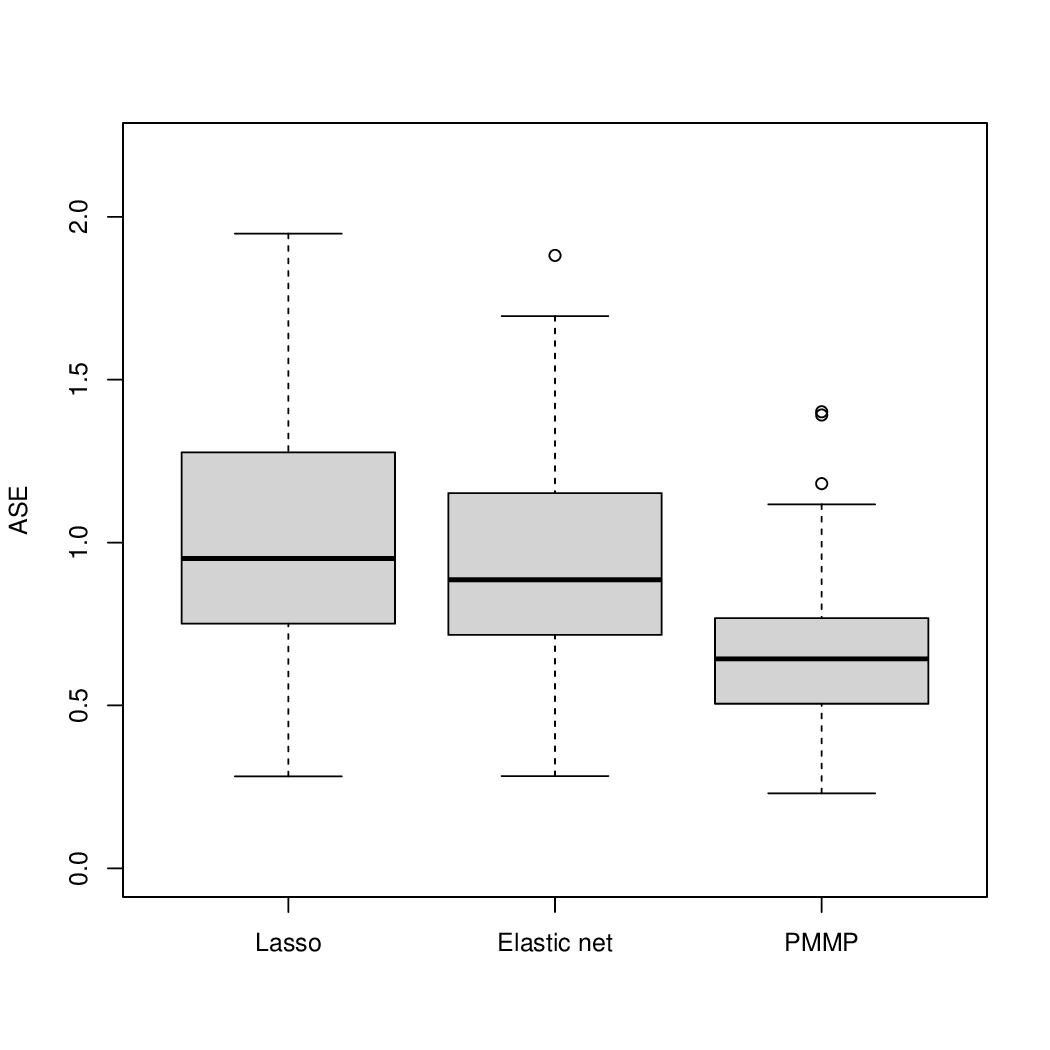}
  \label{2-a}
  }
  \subfigure[$N=50$]{
  \includegraphics[width=0.3\textwidth]{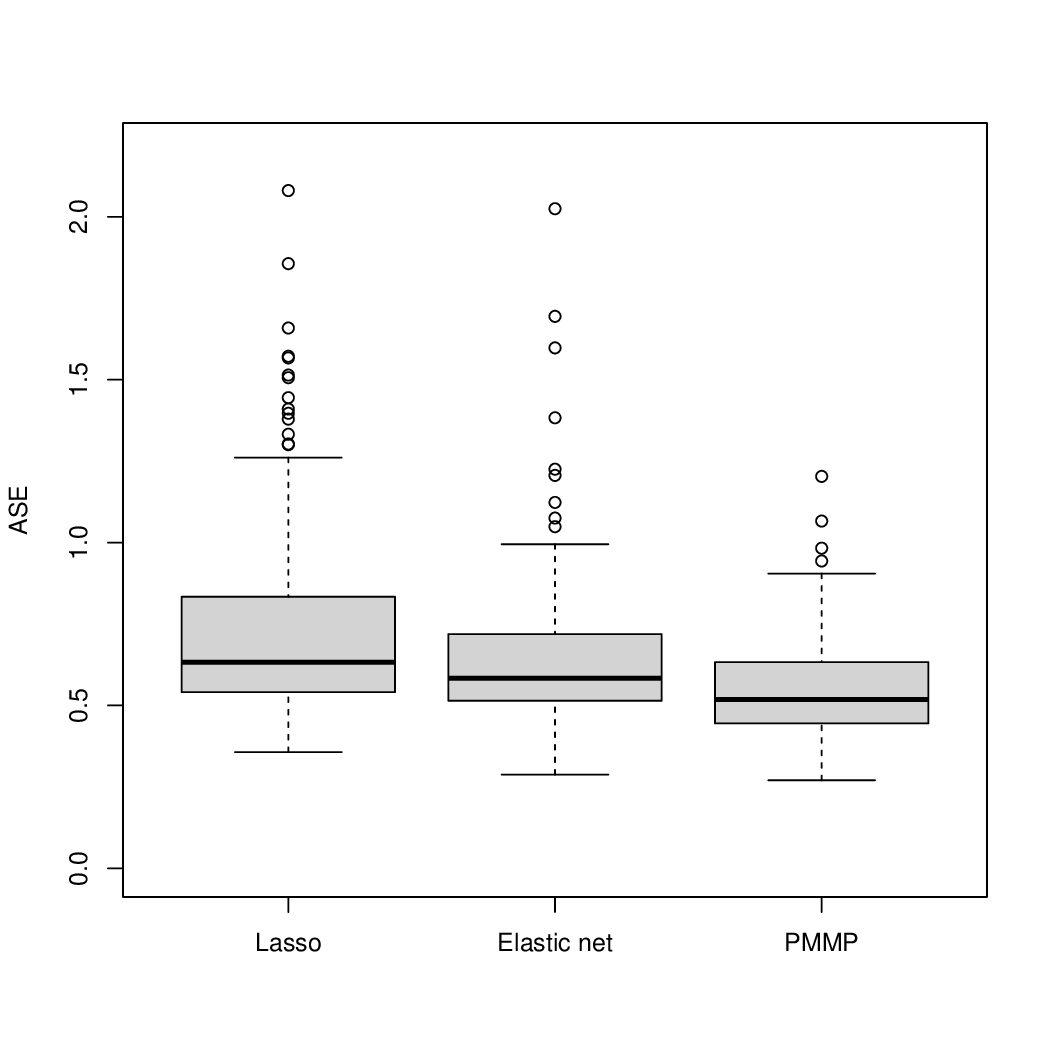}
  \label{2-b}
  }
  \subfigure[$N=100$]{
  \includegraphics[width=0.3\textwidth]{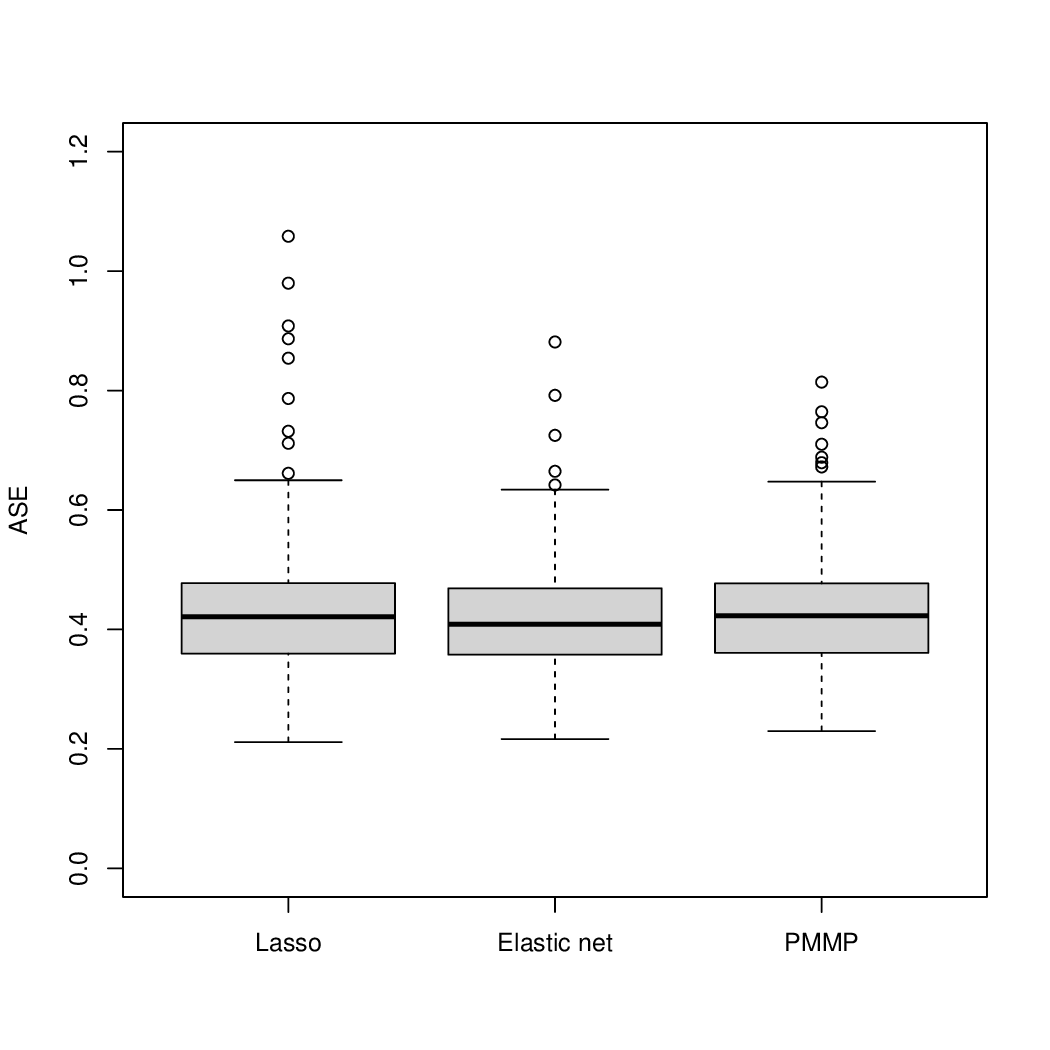}
  \label{2-c}
  }
  \caption{Boxplots of ASEs (Dense Scenario; $N_{\rm sim}=200$)}
\end{figure}

It can be seen that, as $N$ increases, the performance of all three methods improve. When $N$ is much smaller than the
number of predictors (i.e., $N=$ 30 or 50), PMMP seems significantly outperforms the two shrinkage methods. On the
other hand, when $N$ is larger (i.e., $N=100$) such that it is close to the number of predictors (121), the three methods
perform similarly in estimating the regression means, although PMMP still seems to be doing better than the shrinkage
methods in terms of the outliers. Again, note that PMMP does not need to estimate all 121 unknown regression coefficients.

Next, we study the performance of the three methods when $\sigma$ is changing. We consider $N=30$; the other
settings remain unchanged. From Figure 3, it can be seen that as $\sigma$ increases, the performance gap between
different methods reduces, but PMMP still seems to perform better than the other two methods.
\begin{figure}[t!]
  \subfigure[$\sigma=0.8$]{
  \includegraphics[width=0.3\textwidth]{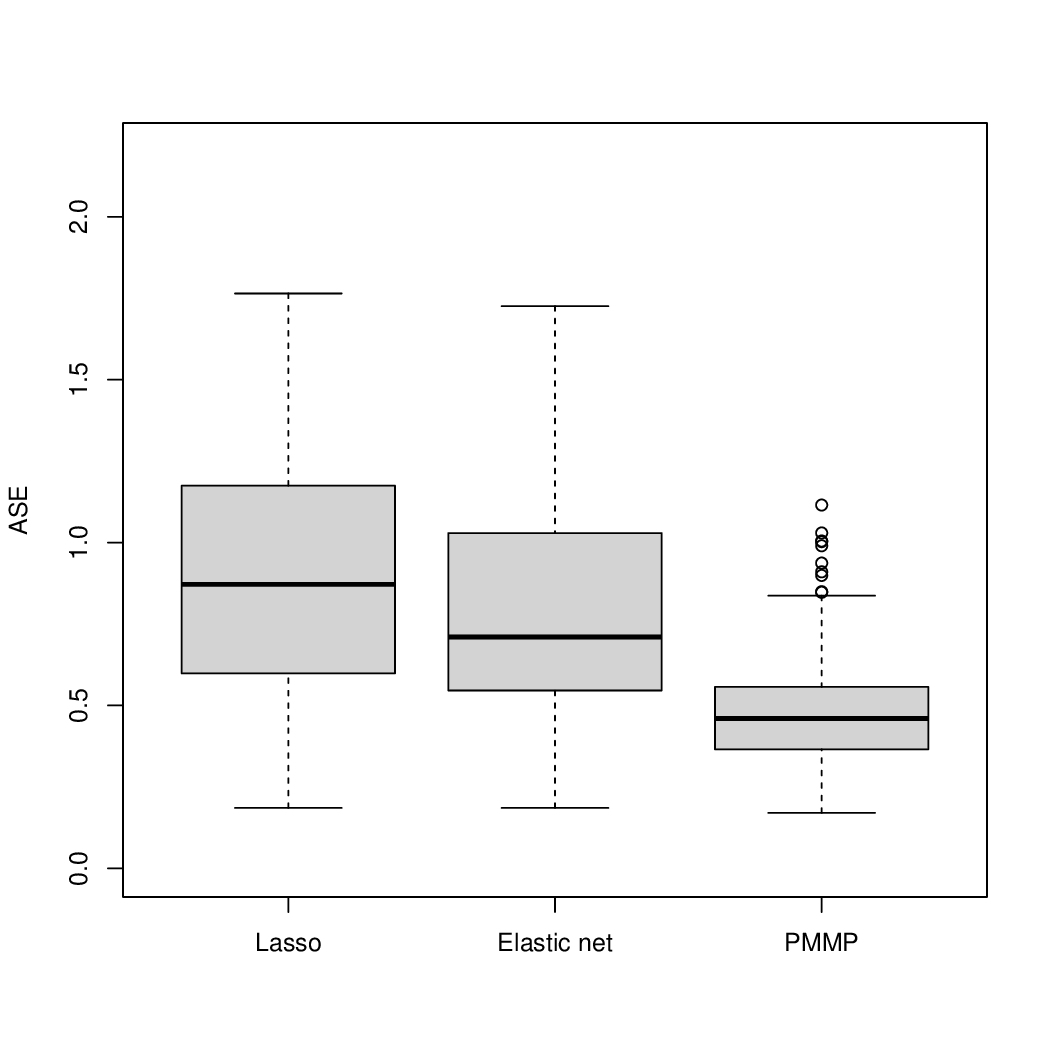}
  \label{3-a}
  }
  \subfigure[$\sigma=1$]{
  \includegraphics[width=0.3\textwidth]{simu-U01-sigma1-30.eps}
  \label{3-b}
  }
  \subfigure[$\sigma=2$]{
  \includegraphics[width=0.3\textwidth]{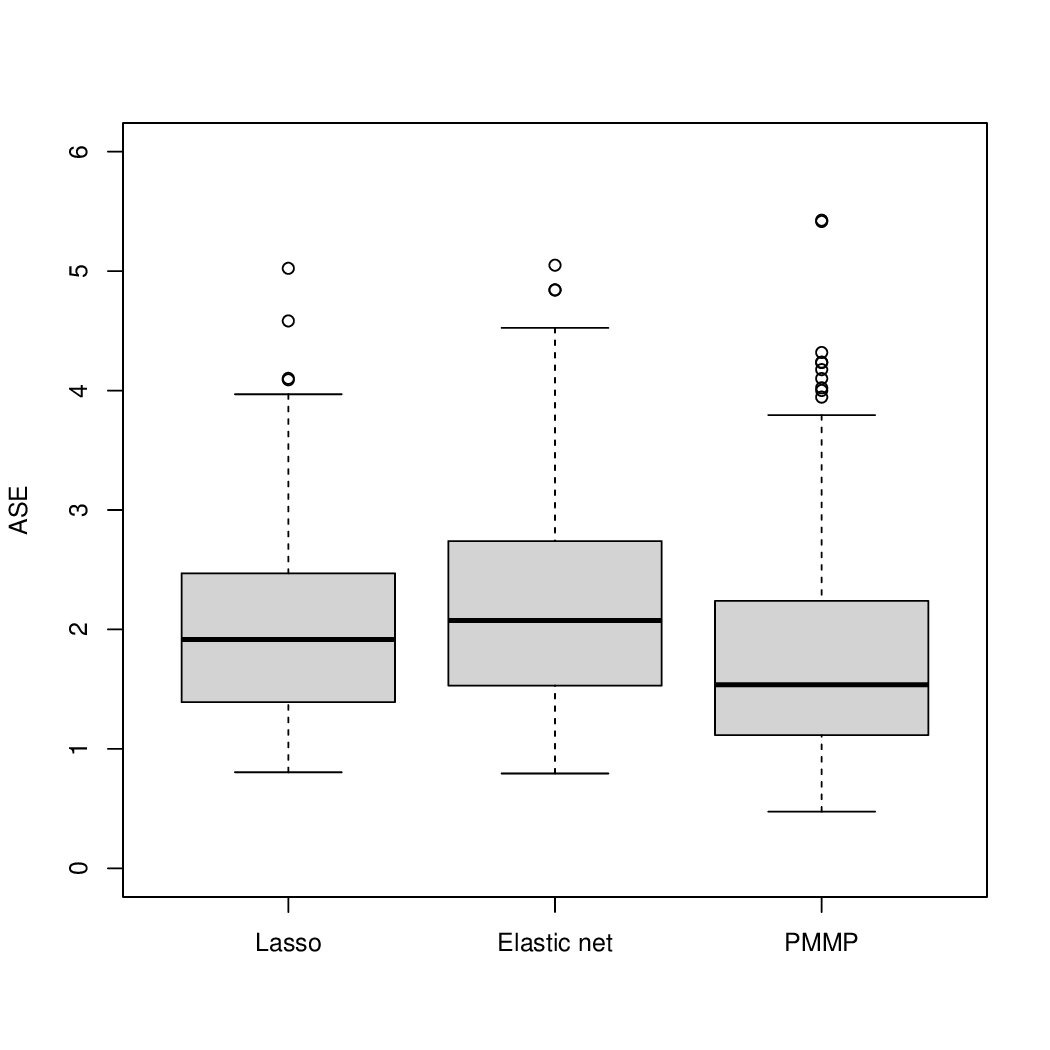}
  \label{3-c}
  }
  \caption{Boxplots of ASEs (Dense Scenario; $N=30$, $N_{\rm sim}=200$)}
\end{figure}

Finally, we compare the three methods under different settings of categorical predictors so that the groups classified by
PMMP are different for the same $N$. We consider the following settings:
(a) Data are generated by model \eqref{eq:sim_model} without the categorical predictor $c_{i1}$; two-way interactions are included; the rest remain the same.
(b) Data are generated by model \eqref{eq:sim_model} without the categorical predictor $c_{i2}$; two-way interactions are included; the rest remain the same.
(c) Data are generated by model \eqref{eq:sim_model} without the categorical predictor $c_{i3}$; two-way interactions are included; the rest remain the same.
(d) Data are generated by model \eqref{eq:sim_model} but with only one categorical predictor $c_{i3}$; the rest remain the
same.
Figure 4 shows that, under the different settings, the number of groups, $K$, classified by PMMP ranges from 6 to 20. The performance of PMMP still seems significantly better than the other two methods.
\begin{figure}[t!]
  \subfigure[Without $c_{i1}, K=19$]{
  \includegraphics[width=0.45\textwidth,height=4cm]{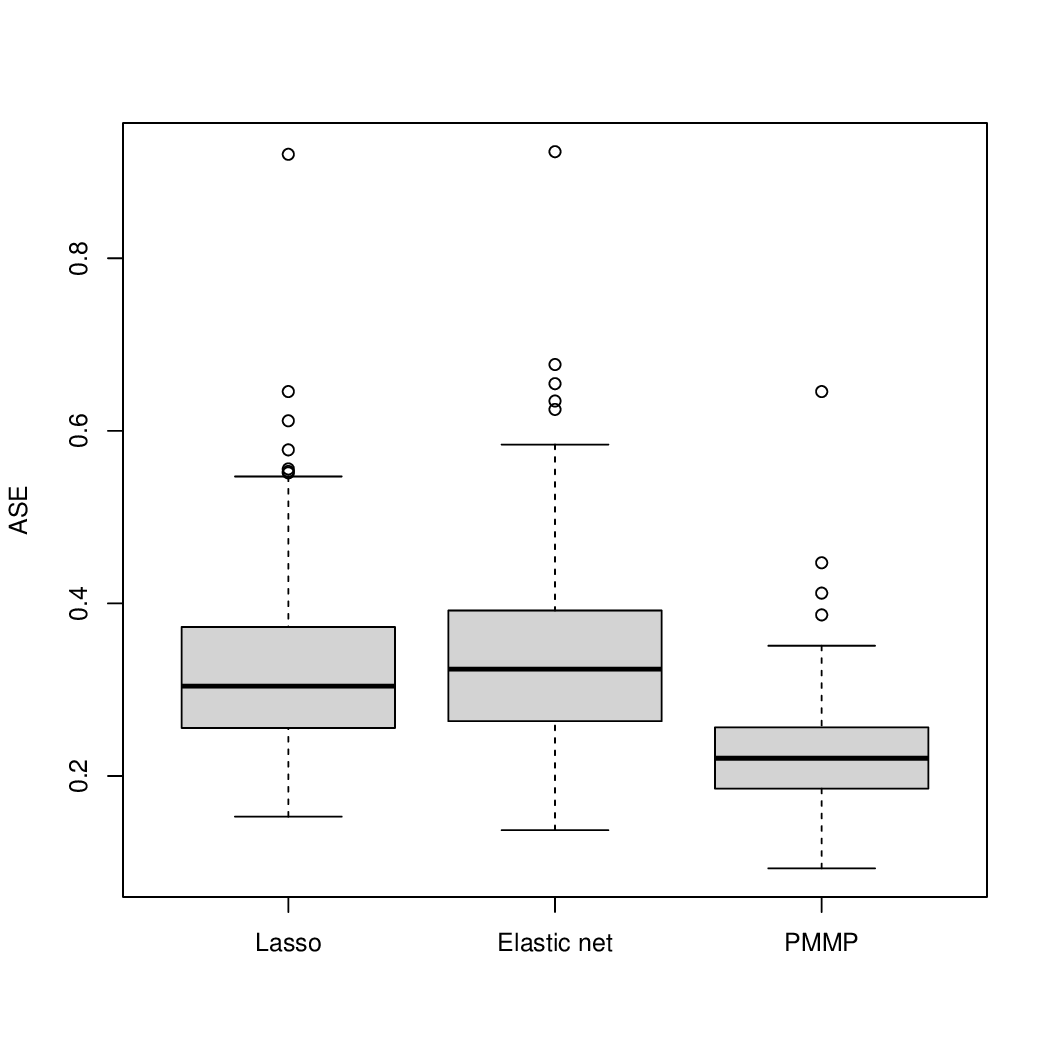}
  \label{4-a}
  }
  \subfigure[Without $c_{i2}, K=20$]{
  \includegraphics[width=0.45\textwidth,height=4cm]{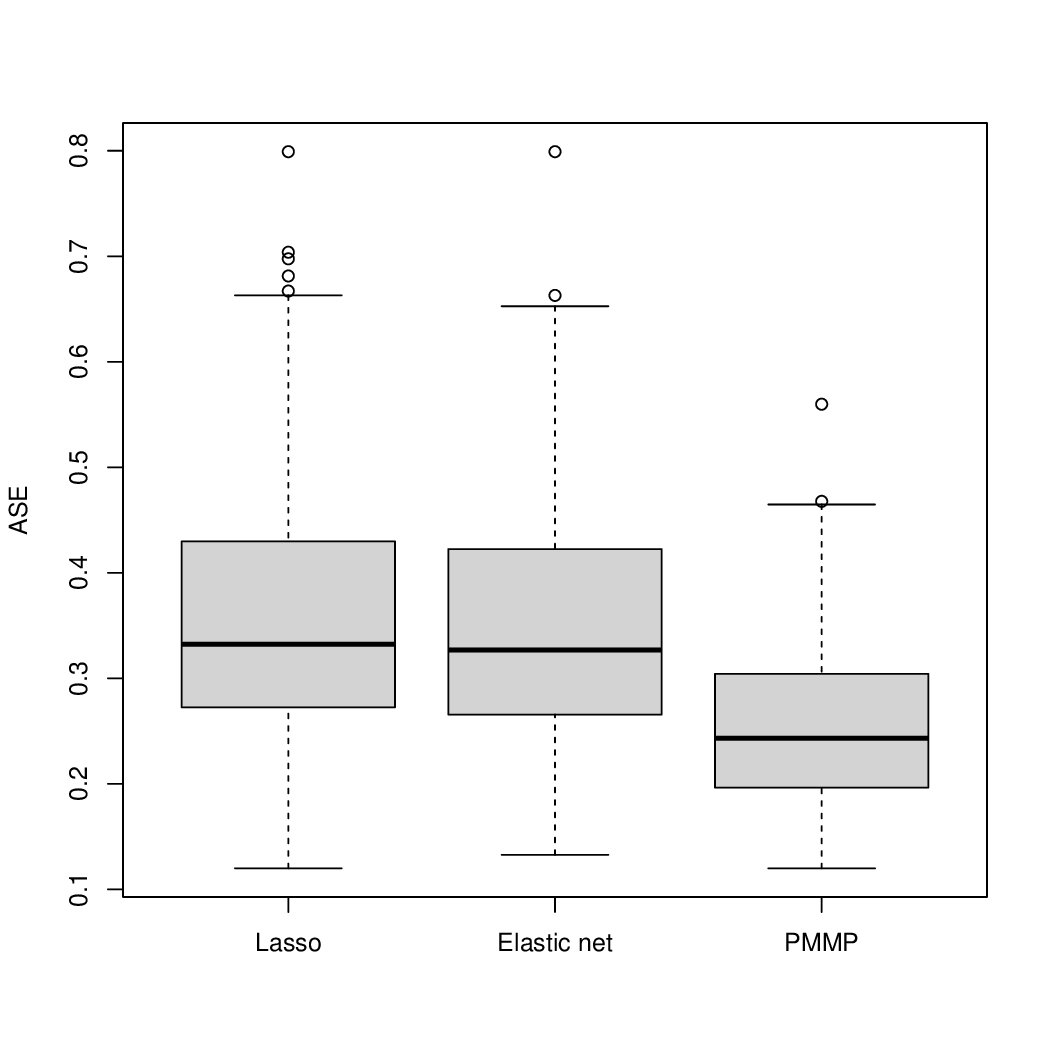}
  \label{4-b}
  }
  \subfigure[Without $c_{i3}, K=16$]{
  \includegraphics[width=0.45\textwidth,height=4cm]{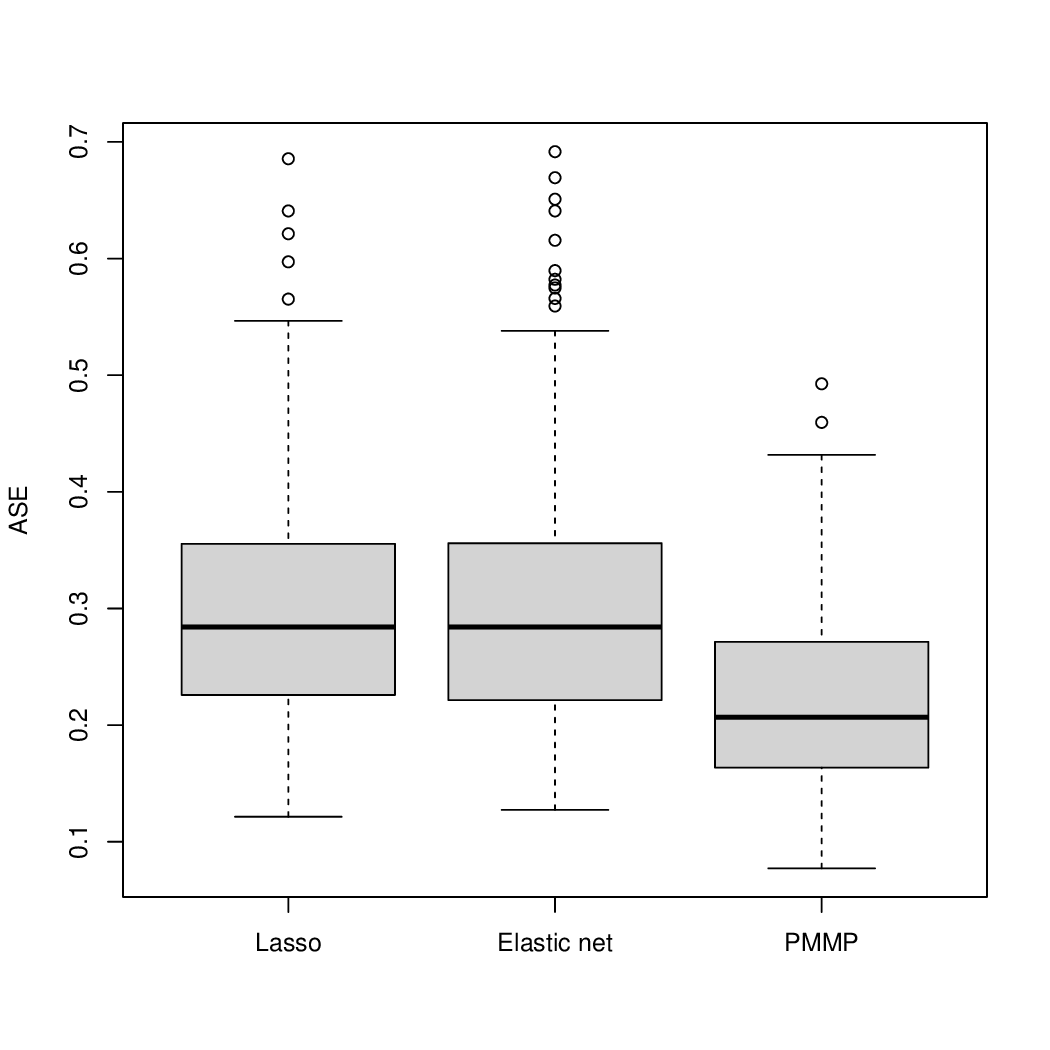}
  \label{4-c}
  }
  \subfigure[Only $c_{i3}, K=6$]{
  \includegraphics[width=0.45\textwidth,height=4cm]{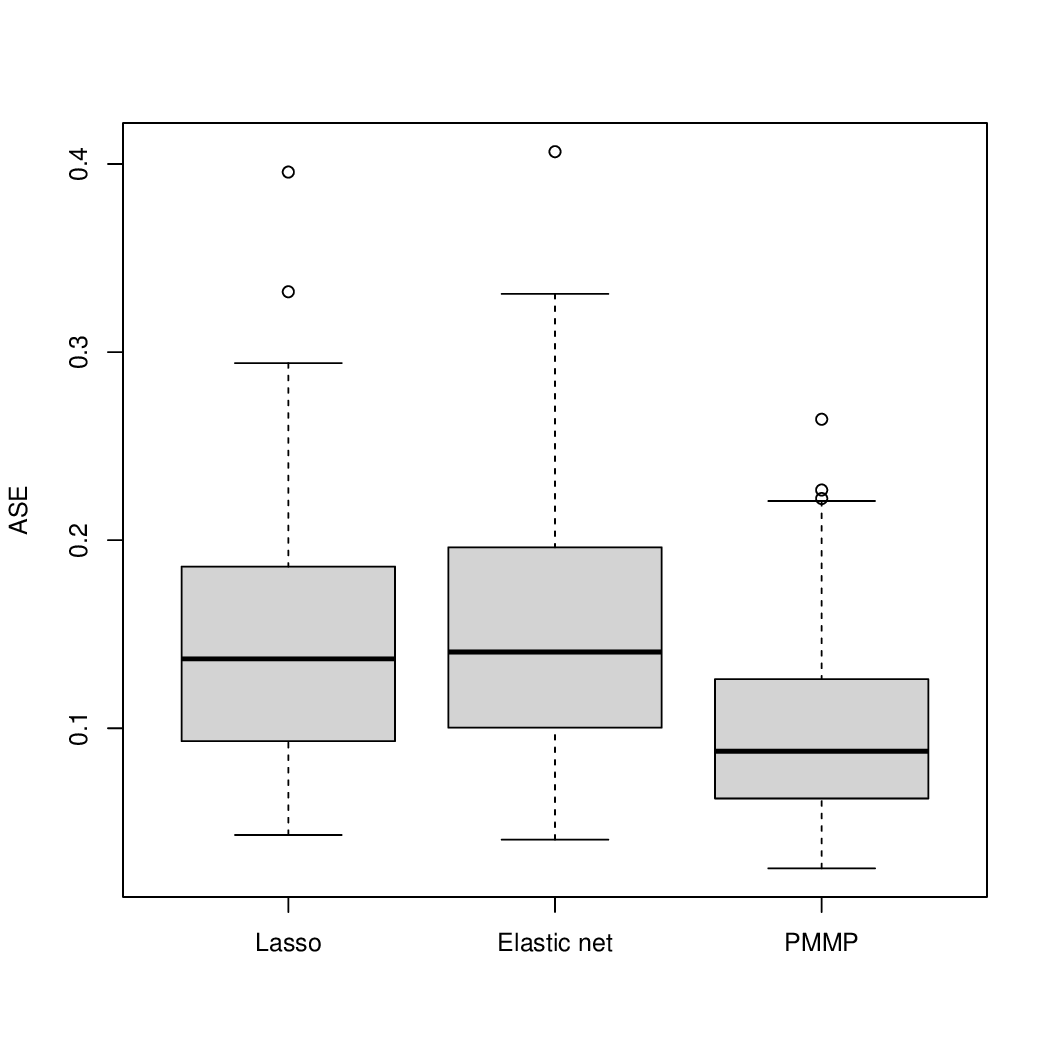}
  \label{4-d}
  }
  \caption{Boxplots of ASEs ($N=50$, $\sigma=1$, $N_{\rm sim}=200$)}
\end{figure}

Additional simulation results are deferred to Supplementary Material.
\subsection{Performance of MSE estimator}
\label{sec:4th-2}
\hspace{4mm}
Under the same simulation setting, we study empirically the performance of the proposed MSE estimator for PMMP. We
consider $N=50, 100$, which were previously considered, and a larger sample size, $N=200$, to see the improvement
of the MSE estimator as the sample size increases.

We increase the number of simulation runs to 1,000 to obtain more accurate results. Under each sample size, $N$,
we evaluate the true MSE based on the simulation runs, that is, by computing
${\rm MSE}_{i}=N_{\rm sim}^{-1}\sum_{s=1}^{N_{\rm sim}}(\hat{\theta}_{i,s}-\theta_{i,s})^{2}$,
where $\theta_{i,s}$ is the true regression mean for the $i$th observation (which is known because that is
how we simulated the data), and $\hat{\theta}_{i,s}$ is the corresponding PMMP, for the $s$th simulation run, $1\leq
s\leq N_{\rm sim}$. We then compute the simulated mean of the MSE estimator, again over the simulation runs, that
is, ${\rm E}(\widehat{\rm MSE}_{i})=N_{\rm sim}^{-1}\sum_{s=1}^{N_{\rm sim}}\widehat{\rm MSE}(\hat{\theta}_{i,s})$,
where $\widehat{\rm MSE}(\hat{\theta}_{i,s})$ is the MSE estimate for $\hat{\theta}_{i,s}$, given at the end of Section
3, for the $s$th simulation run, $1\leq s\leq N_{\rm sim}$. The relative bias (RB) is defined as ${\rm RB}_{i}=\{{\rm
E}(\widehat{\rm MSE}_{i})-{\rm MSE}_{i}\}/{\rm MSE}_{i}=\{{\rm E}(\widehat{\rm MSE}_{i})/{\rm MSE}_{i}\}-1$, for
$1\leq i\leq N$. A boxplot for the $N$ RBs are presented in Figure 5, for $N=50, 100, 200$.

\begin{figure}[t!]
  \subfigure[$N=50$]{
  \includegraphics[width=0.3\textwidth]{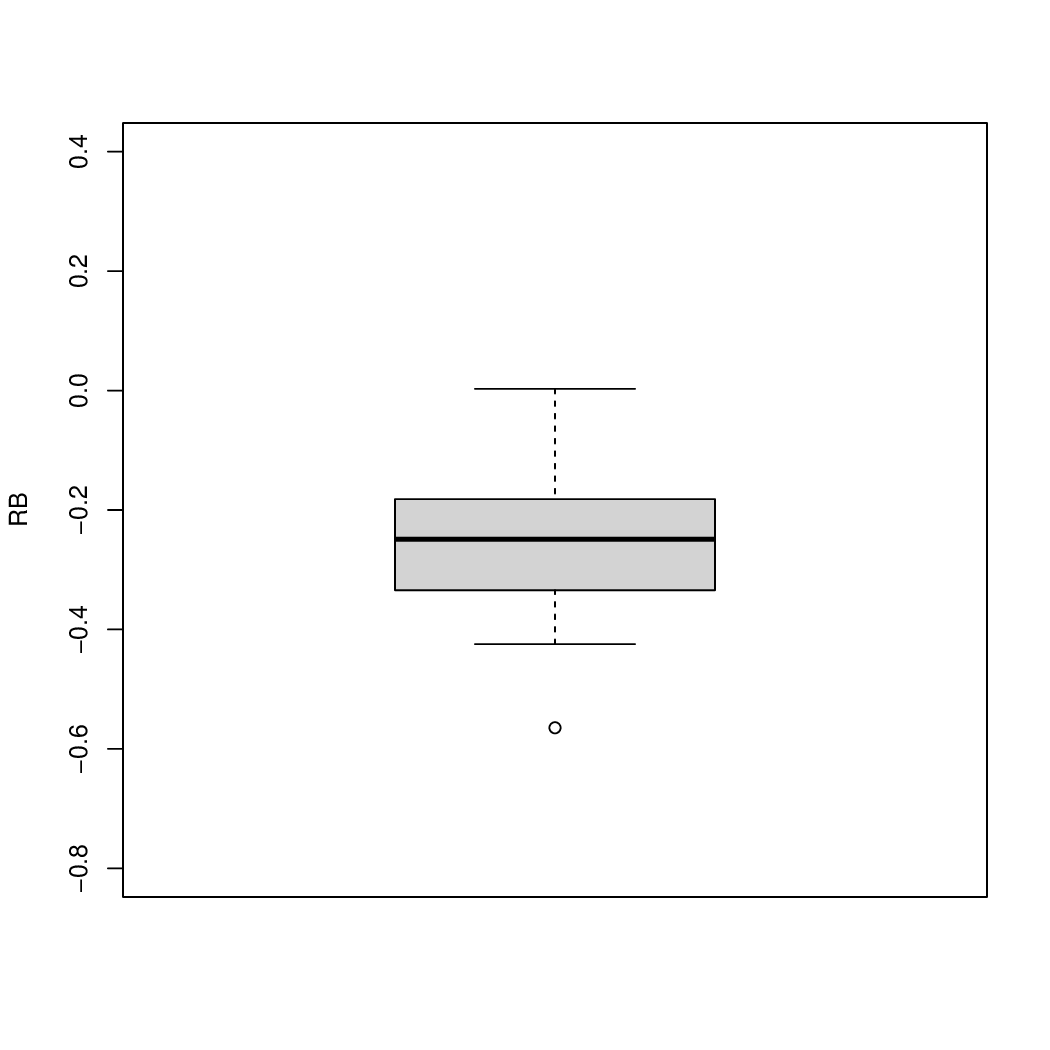}
  \label{5-a}
  }
  \subfigure[$N=100$]{
  \includegraphics[width=0.3\textwidth]{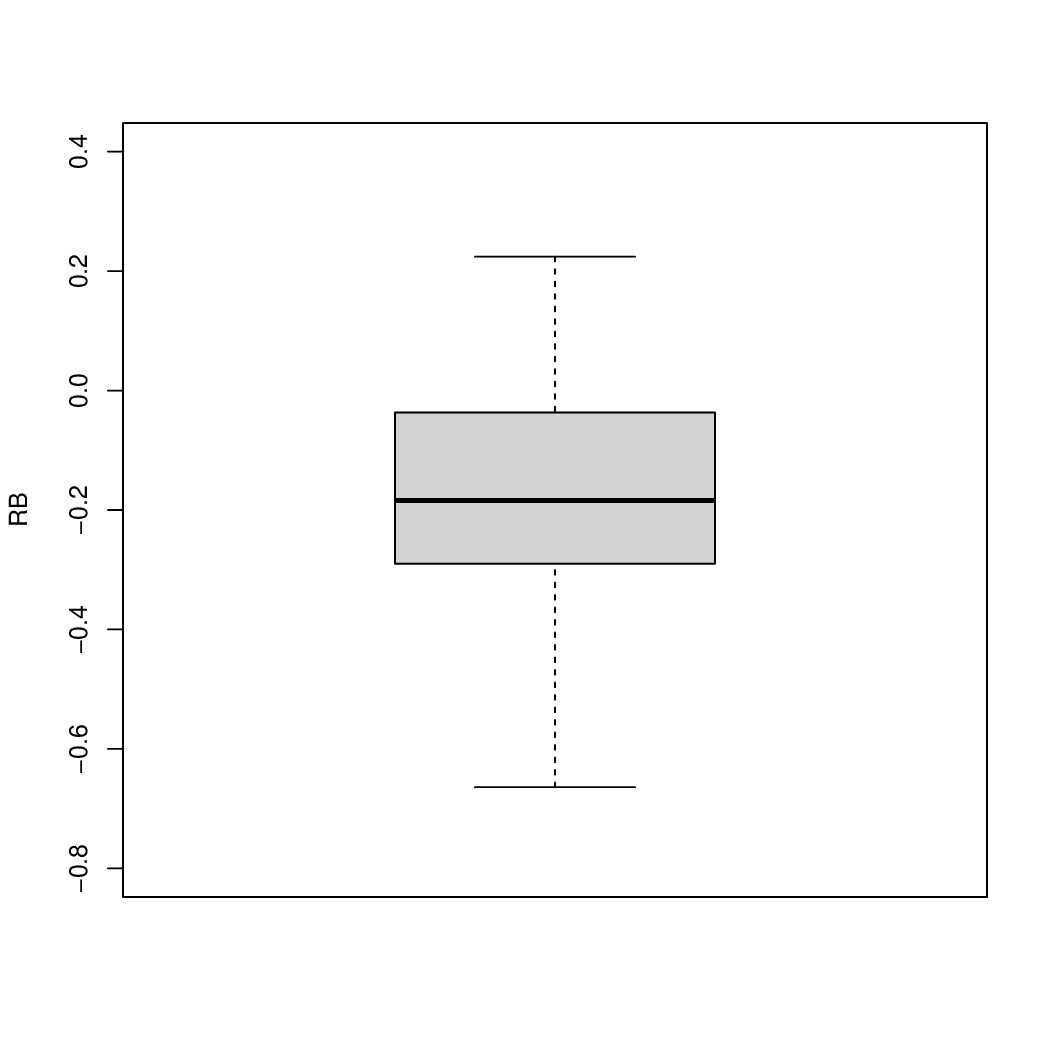}
  \label{5-b}
  }
  \subfigure[$N=200$]{
  \includegraphics[width=0.3\textwidth]{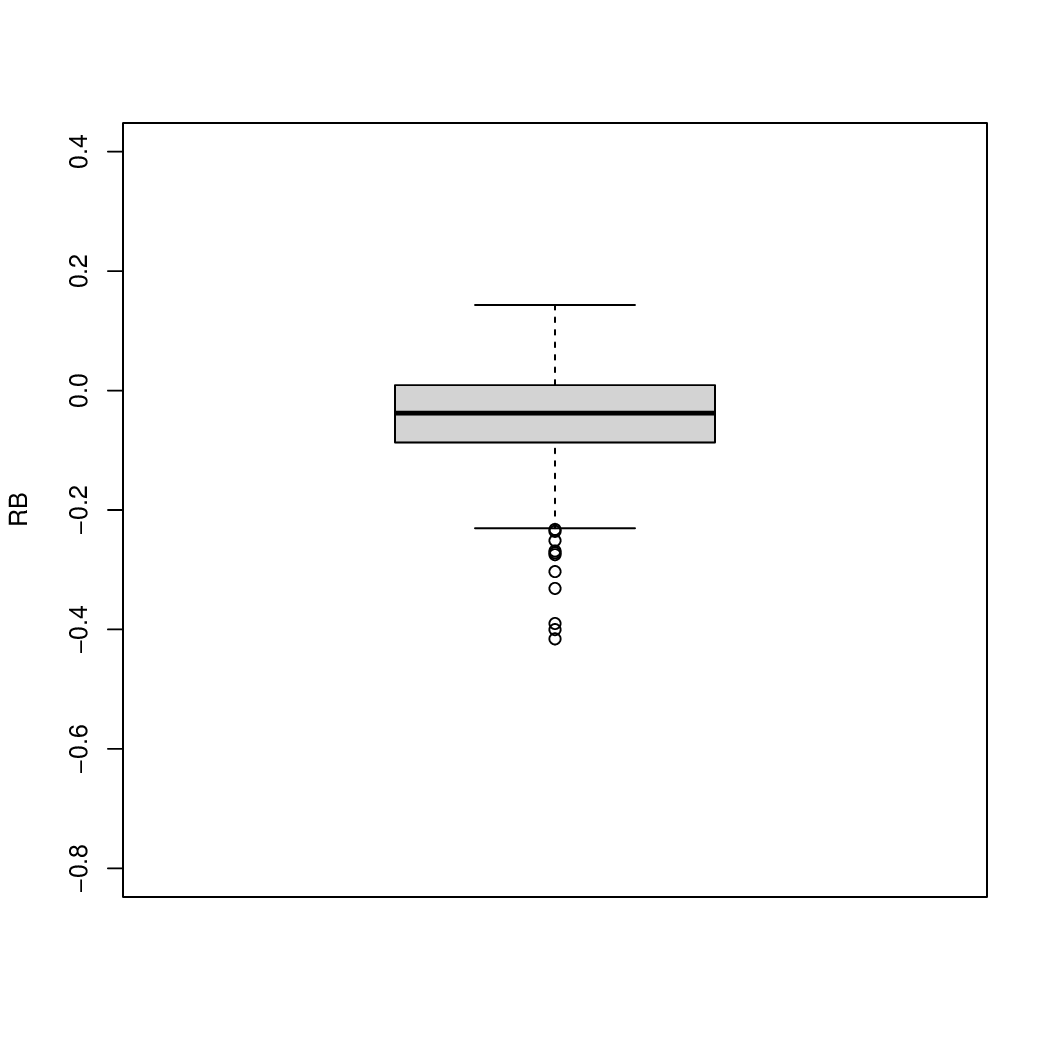}
  \label{5-c}
  }
  \caption{Boxplots of RBs (Dense Scenario; $N_{\rm sim}=1000$)}
\end{figure}

The improvement of the performance of the MSE estimator, as $N$ increases, is evident from the figure. When $N$
is relatively small ($N=50$), there is a negative relative bias, indicating underestimation of the true MSE. The absolute
values of RB are generally (much) less than 0.5, with a median around $-0.23$, and inter-quantile range (IQR)
between $-0.32$ and $-0.18$. For $N=100$, the absolute values of RB are still mostly (much) less than 0.5, with a
median around $-0.19$ and IQR between $-0.29$ and $-0.08$. A more significant improvement is seen with $N=200$,
with all absolute values of RB (much) less than 0.5, median around $-0.04$, and IQR between $-0.09$ and $0.02$.

Also note that, when $N$ increases from 50 to 100, there is a apparent increase in terms of the spread of the RB
values. This is largely due to the fact that more RB values are contributing to the boxplot for $N=100$ than to the
boxplot for $N=50$. However, even this factor is overcome when $N$ further increases, as is apparent in the boxplot
for $N=200$.
\section{Bone marrow data revisited}
\label{sec:5th}
\hspace{4mm}
We use the ``Bone marrow transplant: children Data Set'' in the UCI Machine Learning Repository to illustrate PMMP and compare it with the shrinkage methods. The data set is collected from 187 pediatric patients with 39 attributes. Some
attributes describe similar information, such as donor\_age and donor\_age\_below\_35. Finally, we selected 6 continuous
variables and 8 categorical variables in a regression analysis with the outcome variable being the survival time of patients.
See Table 1 of the supplement for the variable explanations. Among the categorical variables, CMV\_status has 4
categories and HLA\_group\_1 has 7 categories; the rest all have two categories. The main purpose of the analysis is to
estimate the mean survival times. After removing missing values, there are 166 samples left for analysis. All continuous
variables are standardized, as is typical for analyses using the shrinkage methods (see below). The response variable is
log-transformed.

For the Lasso/Elastic net methods, we consider the linear model with all of the selected variables, plus the two-way and
three-way interactions among the categorical variables. The total number of predictors is $6+15+87+263=371$, far
exceeding the sample size $n=166$. For the PMMP method, based on the 8 categorical predictors, the 166 samples are
classified into $K=130$ groups. Note that, unlike the simulation, here we do not know the true values of $\theta_i$, which
is the mean survival time for this real data. Thus, it is not possible to compare the exact performances of the different
methods. Nevertheless, note that the standard regression predicted value for $y_{i}$ (if it were unobserved) is the same
as the estimated mean of $y_{i}$. Thus, we may compare the mean squared prediction error (MSPE) for predicting $y_{i}$
using different methods. The boxplots of the squared prediction errors for the three comparing methods are presented
in Figure 6.

\begin{figure}[t!]
  \centering
  \includegraphics[width=0.7\textwidth,height=6cm]{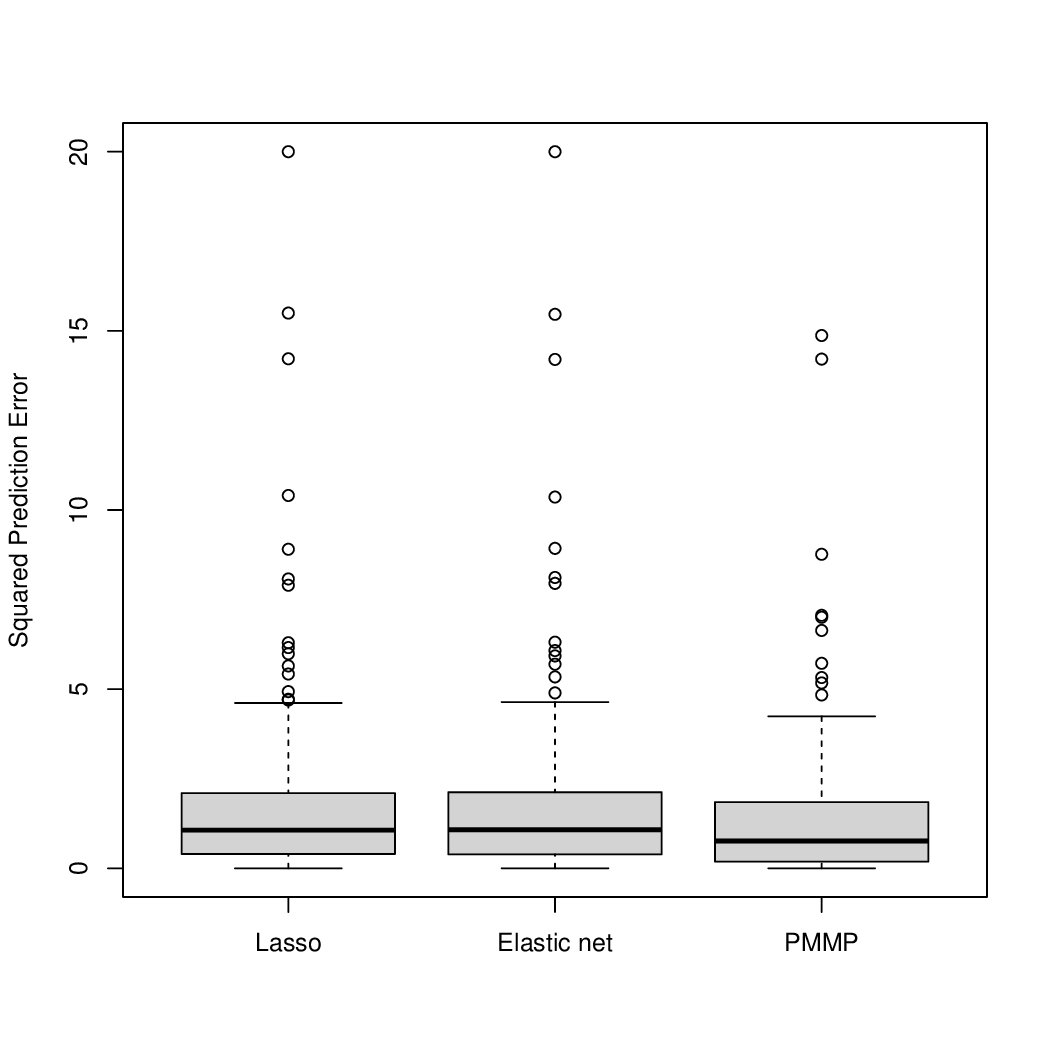}
\caption{Real-data Example: Boxplots of Squared Prediction Errors}
\end{figure}

The figure shows that PMMP is doing moderately better than the two shrinkage methods in terms of the
squared prediction error. Although, as noted this, this is not an accurate evaluation of the performance,
it may, at least, tell us something that is relevant.

Finally, we obtain the MSE estimate for each predicted value (i.e., pseudo EBLUP), $\hat{\theta}_{i}$, then
use 2 times the square root of the MSE estimate as a margin of error. The pseudo EBLUPs (red circles),
with the corresponding margins of error (plus/minus), expressed as the (black) vertical bars centered at the
pseudo EBLUPs, are presented in Figure 7.

\begin{figure}[t!]
  \centering
  \includegraphics[width=0.7\textwidth,height=6cm]{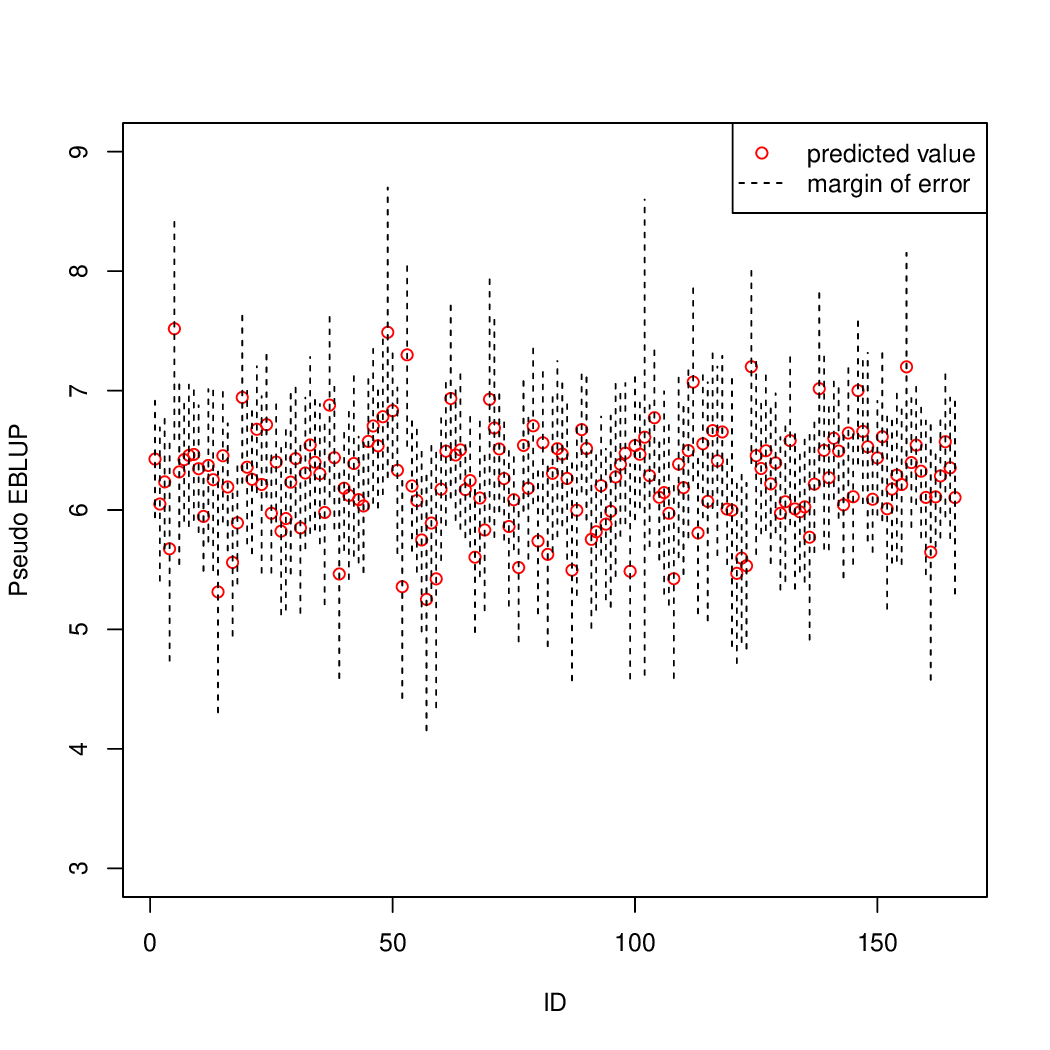}
\caption{Real-data Example: Pseudo EBLUPs with Margins of Error}
\end{figure}

As noted, the main purpose of PMMP is to estimate the regression mean function, rather than interpret the
relationship between the outcome variable and the predictors. Although it is possible to extend the method to
address the interpretation interested, as noted in the first paragraph of Section 2, this requires additional
development on measures of uncertainty for estimating the regression coefficients (of the continuous
predictors or, more generally, any predictors of inferential interest). At the current stage, our method is not
ready to compare with Lasso or elastic net regarding interpretation or variable selection.
\section{Discussion and concluding remark}
\label{sec:6th}
\hspace{4mm}
Although the proposed method is intended for estimation of the mean response, or outcome, a straightforward
extension can be made if one is also interested in interpreting the relationship between the outcome variable
and some of the categorical predictors, or knowing whether some of the categorical predictors are important.
To do so, all one has to do is to separate those categorical predictors, whose relationships with the outcome
are of interest, and include them as part of $x_{i}$. So, in this case, $x_{i}$ includes not only the continuous
predictors but also some categorical predictors of interpretation or inferential interest (see Section 4); the rest
of the categorical predictors are treated the same way as described in Section 2.

Although, in this paper, we have focused on linear models, the basic idea of PMMP can be extended to
generalized linear models (GLM; McCullagh and Nelder 1989), using similar prediction methods developed
in generalized linear mixed models (GLMM); see, for example, Jiang and Nguyen (2021, sec. 3.6). As noted
by the latter authors, the GLMM analogy of EBLUP may be viewed as maximum a posterior estimator.
An existing shrinkage method that applies to GLM is elastic net (Zou and Hastie 2005). Detailed development
in this direction is beyond the scope of this paper.

What is more, the PMMP idea may have a broader implication to high-dimensional statistical inference:
Target the characteristics of direct interest. Sometimes, or often time, such characteristics, altogether, is
of lower dimension than all of the unknown parameter involved in the model. If this is the case, there is
no need to estimate the unknown parameters themselves; rather, one can focus on some functions of
parameters that are of direct interest. PMMP is a testimony of such a simple idea at work.
\section*{Supplementary Materials}

The Supplementary Material provides proofs of Theorem 1 and Theorem 2, the more explicit expressions for $W$ in section 3, a table of variable description, and additional simulation results.

\vspace{5mm}

\section*{Acknowledgements} The research of Hanmei Sun is partially supported by the National Natural Science
Foundation of China (Grant no. 12001334), the Natural Science Foundation of Shandong Province (Grant
no.  ZR2020QA022). The research of Jiangshan Zhang and Jiming Jiang is partially supported by the NSF
grants DMS-1914465 and DMS-2210569.

\end{document}